\documentstyle [prl,aps]{revtex}
\newcommand{\be}{\begin{equation}}
\newcommand{\ee}{\end{equation}}
\newcommand{\bea}{\begin{eqnarray}}
\newcommand{\eea}{\end{eqnarray}}

\newcommand{\p}{\partial}

\newcommand{\up}{\uparrow}
\newcommand{\down}{\downarrow}

\newcommand{\ri}{\mbox{i}}
\newcommand{\re}{\mbox{e}}

\begin {document}
\bibliographystyle {plain}
%\tableofcontents
\begin{titlepage}
\begin{flushright}
\today
\end{flushright}
\vspace{0.5cm}
\begin{center}
{\Large {\bf Chirally Stabilized Critical State in Marginally
Coupled Spin and Doped Systems }}\\ 
\vspace{1.8cm}
\vspace{0.5cm}
{P. Azaria$~^{1}$ and P. Lecheminant$~^{2}$}\\
\vspace{0.5cm}
{\em$^{1}$
Laboratoire de Physique Th\'eorique des Liquides\\
Universit\'e Pierre et Marie Curie, 4 Place Jussieu, 75252 Paris, France}\\
{\em$^{2}$ Laboratoire de Physique Th\'eorique et Mod\'elisation,\\
Universit\'e de Cergy Pontoise, 5 Mail Gay-Lussac, Neuville sur Oise, 
95301 Cergy
Pontoise Cedex, France}

\begin{abstract}
\par
We study a class of one-dimensional models consisting of 
a frustrated (N+1)-leg spin ladder, its asymmetric doped version
as a special example of a Luttinger liquid in an active environment,
and the N-channel Kondo-Heisenberg model away from half-filling.
It is shown that these models exhibit a critical phase with 
generally a non-integer central charge and belong to 
the class of chirally stabilized spin liquids recently
introduced by Andrei, Douglas, and Jerez [Phys. Rev. B 58, 7619 (1998)].
By allowing anisotropic interactions in spin space, 
an exact solution in the N=2 case is found 
at a Toulouse point which captures all
universal properties of the models.
At the critical point, the massless degrees of freedom 
are described in terms of an effective S=1/2 Heisenberg spin chain 
and two critical Ising models.
The Toulouse limit solution enables us to discuss the 
spectral properties, the 
computation of the spin-spin correlation functions as well as 
the estimation of the NMR relaxation rate
of the frustrated three-leg ladder. Finally, it is shown
that the critical point becomes unstable upon switching on 
some weak backscattering perturbations in the frustrated three-leg ladder.
\end{abstract}
\end{center}

\end{titlepage}
\sloppy
\par
\section{Introduction}

One-dimensional quantum spin systems exhibit fascinating physical 
properties. Most of the striking features obsverved in these 
systems are pure quantum effects and give rise to many 
exotic ground-states or unconventional excitation spectra that 
are not accessible by traditional approaches like spin-wave or 
perturbation theory.
A famous example is the prediction made by Haldane\cite{haldane} that 
half-integer antiferromagnetic spin chains are gapless whereas 
those with integer spin have a finite energy gap in its spectrum.

The different one-dimensional spin gapped phases are usually
characterized by the nature of the broken discrete symmetries (lattice
translational symmetry for instance) in the ground-state, 
the quantum numbers associated with the massive excitations, and the 
structure of the dynamical magnetic susceptibility.
Many different spin gapped phases have been identified.
The spin-1 Heisenberg chain is certainly the most famous example
of a spin liquid with a gap.
It is characterized by a hidden topological order\cite{den}
related to the breakdown of a hidden Z$_2 \times$ Z$_2$ symmetry\cite{kohmoto}.
The spin excitations are optical magnons carrying spin S=1 and display
a single peak near $q=\pi$ in the dynamical magnetic susceptibility.
The low-lying excitations of a massive phase 
can also have fractional quantum numbers
as in the J$_1$-J$_2$ S=1/2 Heisenberg chain for a sufficiently
large value of J$_2$: The ground-state is 
spontaneously dimerized with a two-fold degeneracy and the fundamental 
excitations are massive S=1/2 solitons (spinons) separating the 
two ground-states\cite{haldane1,sorensen}. 
Another scenario can also be realized in a two-leg spin ladder coupled
by a four spin exchange interaction where the elementary excitation 
is neither an optical magnon nor a massive spinon but rather a pair of 
propagating triplet or singlet solitons connecting the two spontaneously
dimerized ground-states\cite{nersesyan,kolezhuk}. 
This special structure of the excitation will reveal itself in 
the dynamical magnetic susceptibility which displays a two-particle
threshold instead of a sharp single magnon peak near $q=\pi$ 
as in the S=1 spin chain.
This incoherent background in the dynamical structure factor gives 
rise to a new one-dimensional spin liquid state: the so-called 
non-Haldane spin liquid\cite{nersesyan}.

Clearly, the general classification of all possible one-dimensional
spin gapped phases is still lacking.
However, the integrability of some special field theoretical models 
has given a lot of insight and has been employed to extract 
the degeneracies, quantum numbers of low-energy excited states of spin 
liquid phases and to provide sometimes the computation of the leading 
asymptotics of correlation functions.
This program has been realised for the sine-Gordon model associated
with the low-energy description of the S=1/2 Heisenberg chain with 
alternating exchange\cite{essler}
and the Cu benzoate\cite{oshikawa,essler1}
and also for the O(N) (N=4,6,8) Gross-Neveu model for the description 
of a special frustrated two-leg ladder\cite{allen1}, weakly two-leg 
Hubbard ladder\cite{lin,konik,schulz}, (N,N) armchair
Carbon nanotubes\cite{konik}, and spin-orbital model\cite{azaria}.

The situation is totally different for critical one-dimensional quantum  
spin systems like the spin-1/2 Heisenberg chain where Conformal Field 
Theory (CFT) allows a complete classification of one-dimensional 
quantum systems or finite temperature two-dimensional classical
systems\cite{polyakov,dms}.
The low-energy spectrum of the lattice model with a continuous
symmetry is described in terms of 
representations of a certain current algebra\cite{knizhnik}.
This affine symmetry determines the operator content of the theory
and all possible scaling dimensions of the operators are in turn 
fixed by the conformal invariance of the underlying 
Wess-Zumino-Novikov-Witten (WZNW)
model built from the currents of the affine symmetry
with the Sugawara construction\cite{knizhnik}. 
The knowledge of the scaling dimensions allows the computation of 
correlation functions and the scaling behaviour of the energies of
the ground-state and low-lying states as a function of the size of 
the system\cite{cardy1}.
The different critical phases are labelled, in particular, by
the central charge (c) of the underlying Virasoro algebra of the 
WZNW CFT which fixes the low-temperature behaviour of the specific heat of 
the model\cite{cardy}.

A simple example of this description is the Luttinger 
liquids, a general class of one-dimensional interacting 
electron models whose infrared (IR) behaviour is governed
by the Luttinger model\cite{haldanelutt}, which have an U(1) affine
Kac Moody (KM) symmetry and a central charge c=1.
The Luttinger liquids are usually described in a bosonized form
with the introduction of one gapless bosonic field (c=1 CFT)
and are characterized by single-particle correlation functions having
branch cuts with non-universal exponents\cite{voit}. 
Such state provides an example of non-Fermi liquid behaviour\cite{voit}.
The critical theory of quantum spin systems with non-Abelian continuous
symmetries can also be analysed on the basis of their SU(2)$_k$ affine KM
algebra\cite{affleck0,affleck}.
Examples of such representation are the critical theory of 
half-integer spin antiferromagnets described by the SU(2)$_1$ WZNW model
whereas the family of integrable Hamiltonian of arbitrary spin S\cite{kulish}
belongs to the SU(2)$_{2S}$ universality class\cite{affleck}.
This approach has also been employed to describe the low-energy physics
of Heisenberg antiferromagnets with a larger symmetry group like a SU(N) 
symmetry\cite{sutherland} on the basis of the SU(N)$_1$ WZNW 
model\cite{affleck,affleck1,itoi,assaraf}.

In a recent paper, Andrei, Douglas, and Jerez\cite{andrei} identified 
a new non-Fermi-liquid class of fixed points, the so-called chiral 
spin fluids, describing the IR behaviour of one-dimensional interacting
{\sl chiral} fermions.
The chiral spin liquid (CSL) behaviour appears as a stable critical state of 
one-dimensional interacting fermions with {\sl unequal} numbers of right-moving
and left-moving particles.
The phenomenological Hamiltonian, in the 
continuum limit, corresponding to this state reads:
\bea 
{\cal H}_{CSL} = -\ri v_F \left(\sum_{r=1}^{f_R}
R_{\alpha r}^{\dagger} \partial_x R_{\alpha r} - 
\sum_{l=1}^{f_L}
L_{\alpha l}^{\dagger} \partial_x L_{\alpha l}\right)
+ g {\bf J}_R \cdot {\bf J}_L 
\label{handrei}
\eea
where $R_{\alpha r}^{\dagger}$ (respectively $L_{\alpha l}^{\dagger}$)
is the creation operator of a right-moving (respectively left-moving)
fermion with spin index $\alpha=\up,\down$ and flavor index $r$ (respectively 
$l$): $r=1,..,f_R$ (respectively $l=1,..,f_L$).
The interaction affects only the spin degrees of freedom and consists
of a marginal relevant ($g >0$) current-current term where the spin currents
write in terms of the fermions as follows:
\bea 
J_R^a &=& \frac{1}{2} \sum_{r=1}^{f_R} 
R_{\alpha r}^{\dagger} \sigma_{\alpha \beta}^a R_{\beta r} \nonumber \\
J_L^a &=& \frac{1}{2} \sum_{l=1}^{f_L} 
L_{\alpha l}^{\dagger} \sigma_{\alpha \beta}^a L_{\beta l}, 
\label{currand}
\eea
$\sigma^a$ (a=x,y,z) being the Pauli matrices
and $J_R^a$ (respectively $J_L^a$) belongs to 
the SU(2)$_{f_R}$ (respectively SU(2)$_{f_L}$) KM algebra.
The Hamiltonian (\ref{handrei}), in the spin sector, describes thus
two SU(2)$_{f_R}$ and SU(2)$_{f_L}$ WZNW models 
marginally coupled by a current-current interaction.
For $f_R \ne f_L$, i.e. breakdown of the $R\rightarrow L$ symmetry 
implying breakdown of the time-reversal symmetry, the 
Hamiltonian (\ref{handrei}) belongs to a new class of problems not 
encountered in the above mentionned problems of one-dimensional 
physics.  
By a combination of CFT argument and Bethe Ansatz techniques, 
the authors of Ref. \cite{andrei} showed that the Hamiltonian (\ref{handrei}) 
for $f_R \ne f_L$ flows in the IR limit to an intermediate fixed point
(chiral fixed point) characterizing by universal exponents in the leading
asymptotic of the electronic Green's functions.
From the electronic point of view, such a state provides a non-trivial
example of a non-Fermi-liquid state which differs from 
the Luttinger liquids: the CSL state.
A typical example of possible realizations 
of CSL, given in Ref. \cite{andrei}, is a contact between the edges of 
two integer Quantum Hall Effect systems with different filing factors: 
$\nu_L = f_L \ne \nu_R = f_R$.

In this paper, we shall consider another route
where the total Hamiltonian ($\cal H$) of 
a possible realization of CSL is time-reversal invariant.
Indeed, in the continuum limit, 
the leading part of the Hamiltonian that 
imposes the strong coupling 
behaviour decomposes into two commuting marginal 
{\sl chirally asymmetric} parts ${\cal H}_1$ and ${\cal H}_2$: 
\be
{\cal H} = {\cal H}_1 + {\cal H}_2, \; \; [{\cal H}_1, {\cal H}_2] = 0,
\label{hgenmod}
\ee
where ${\cal H}_1$ and ${\cal H}_2$ are not separately time-reversal 
invariant but they transform into each other under the symmetry $t \rightarrow
-t$.
The model (\ref{hgenmod}) as a whole will display a CSL behaviour in the 
IR limit.
Such scenario suggests that other realizations of this state
can be found in ladder problems.
However,
in the continuum limit of spin ladders, one has, apart from 
marginal current-current interactions, backscattering terms 
${\bf n}_a \cdot {\bf n}_b$ (${\bf n}_a$ being the staggered magnetizations 
of a chain of index a) which have scaling dimension 1 at 
the ultraviolet (UV) fixed point and thus represent strongly 
relevant perturbations.
Therefore, one has clearly to kill this contribution to
have a marginal perturbation and thus to find some specific
realizations of the CSL state. 
We shall follow, in this paper, two different routes to find some 
CSL behaviour in the IR limit of ladder problems.
On the one hand, frustration may play its role either by suppressing 
geometrically
the backscattering contribution 
as in zigzag 
ladders\cite{white,allen,nersesyan1,cabra} or
by allowing several independent coupling constants and 
the possibility to eliminate the unwanted term by a fine
tuning mechanism\cite{azaria1,allen1}. 
In that case, as we shall see later, the CSL found at this 
special point will be destabilized upon switching on 
weak backscattering contributions but there will exist 
an intermediate low-energy region governed by the CSL behaviour.
On the other hand, doping effects can also be useful to 
suppress the backscattering term by some incommensurate effects:
In the continuum limit, 
the unwanted contribution will become a strongly oscillating 
perturbation  that can be ignored in the long distance limit. 

The remainder of the paper is organized as follows:
Different possible realizations of the CSL state
are introduced in Section II:  
the (N+1)-chain cylinder model, its asymmetric doped 
version, and the N-channel Kondo-Heisenberg model 
away from half-filling.
The nature of the IR fixed point (chiral fixed 
point) for these models is determined in 
Section III.
In Section IV, we present a Toulouse point solution in the $N=2$ case
that captures all universal properties of the chiral fixed point.
In that case, the massless degrees of freedom are described 
in terms of an effective S=1/2 Heisenberg spin chain and two critical 
Ising models.
A brief summary of this approach has already been published\cite{azaria1}.
We shall provide here and in Section V
the technical 
details of the Toulouse point solution and in particular 
the determination of the physical properties of the 3-chain cylinder
model.
The nature of the phase diagram of the doped models in the N=2 case
will be addressed in a forthcoming publication\cite{azarial}.
In section VI, the stability of the chiral fixed point of the 
3-chain cylinder model is analysed under the presence of 
weak backscattering perturbations.
Finally, our concluding remarks are 
presented in Section VII. Three Appendixes (A,B,C) give further technical
details necessary for the computation of the leading asymptotics
of the correlation functions occuring in Section V
whereas in the last one the stability of the Toulouse limit solution is
studied.

\section{Possible realizations of chiral spin liquid behaviour}

In this section, we shall present some lattice models that 
are described in the continuum limit by an Hamiltonian with
only marginal current-current interactions.
In the next section, we shall show that the low-energy physics of this
Hamiltonian is governed by the chiral fixed point and thus represents
some specific realizations of a CSL state.

\subsection{The (N+1)-chain cylinder model}

The first model that we consider below can be visualized 
as follows. It consists of N identical S=1/2 antiferromagnetic
Heisenberg chains on the surface of a cylinder parallel to each other
and to the axis of the cylinder.
An additional S=1/2 Heisenberg chain, that will be called 
the zeroth chain or central chain in the following, resides 
inside the cylinder along
its axis. The surface chains do not interact with each other but 
all of them are coupled to the central one by on-rung interchain
interaction ($J_{\perp}$) and an interchain interaction along 
diagonals of the elementary plaquettes ($J_{\times}$).
The Hamiltonian corresponding to this model is given by:
\be
{\cal H} = \sum_{a=0}^{N} {\cal H}_a^{0} + {\cal H}_{int},
\label{hamiltmod1}
\ee
where ${\cal H}_a^{0}$
is the Hamiltonian of a S=1/2 antiferromagnetic 
Heisenberg spin chain of index a: 
\be 
{\cal H}_a^{0} = J_{\parallel} \sum_j {\bf S}_{a,j}
\cdot {\bf S}_{a,j+1}, \; \; J_{\parallel} > 0
\ee
${\bf S}_{a,j}$ being S=1/2 spin operators on each site $j$ of the
chain of index $a=0,1,..,N$.
These Heisenberg chains are coupled with the interaction:
\bea 
{\cal H}_{int} = 
J_{\perp}  \sum_j \sum_{a=1}^{N} {\bf S}_{0,j}\cdot {\bf S}_{a,j}
+ J_{\times} \sum_j
\sum_{a=1}^{N} \left[{\bf S}_{a,j}
\cdot {\bf S}_{0,j+1} + {\bf S}_{a,j+1} \cdot {\bf S}_{0,j}\right].
\label{hamilintmod1}
\eea
In the following, it will be assumed that all interchain
exchange interactions 
are positive and weak: $0 < J_{\perp},J_{\times} \ll J_{\parallel}$
to describe the model by a continuum limit approach.
Due to the presence of the plaquette interaction $J_{\times}$,
the cylinder model belongs to the class of frustrated spin 
ladders. For $N=1$, this model has been investigated by 
several groups\cite{oitmaa,allen1} and reveals no CSL state.
As it will be discussed in Section III, the (N+1)-chain cylinder model
with $N>1$ is a better candidate for displaying non-trivial CSL 
physics.

Let us first consider the continuum limit of the (N+1)-chain cylinder model.
The continuum description of the S=1/2 antiferromagnetic
Heisenberg spin chain is based on the Sugawara representation
of the SU(2)$_1$ WZNW model
with a marginal irrelevant perturbation\cite{affleck0,affleck}:
\be
{\cal H} ^0_a = \frac{2\pi v}{3} \; \left({\bf J}_{aR} \cdot {\bf J}_{aR} +
{\bf J}_{aL} \cdot {\bf J}_{aL} \right) +
\gamma {\bf J}_{aR} \cdot {\bf J}_{aL}, \; \; a=0,1,..,N
\label{hcin0}
\ee
with $\gamma < 0$ and 
${\bf J}_{aR,L}$ are the right and left spin currents
belonging to the SU(2)$_1$ KM algebra\cite{knizhnik,dms,book,book1} 
and verify the following operator product expansion (OPE):
\bea
J_{aL}^{\alpha}\left(z\right) J_{bL}^{\beta}\left(w\right) &\sim&
\frac{\ri \epsilon^{\alpha \beta \gamma} \delta_{ab}}{2\pi \left(z-w\right)}
J_{aL}^{\gamma}\left(w\right)
+ \frac{\delta^{\alpha \beta}\delta_{ab}}{8\pi^2\left(z-w\right)^2} 
\nonumber \\
J_{aR}^{\alpha}\left(\bar z\right) J_{bR}^{\beta}\left(\bar w\right) &\sim&
\frac{\ri \epsilon^{\alpha \beta \gamma}\delta_{ab}}{2\pi
\left(\bar z- \bar w\right)}
J_{aR}^{\gamma}\left(\bar w\right)
+ \frac{\delta^{\alpha \beta}\delta_{ab}}{8\pi^2
\left(\bar z- \bar w\right)^2} \nonumber \\
J_{aL}^{\alpha}\left(z\right)
J_{bR}^{\beta}\left(\bar w\right) &\sim& 0, \; \; \alpha,\beta=x,y,z, \; \; 
a,b=0,1,..,N 
\label{opesu2level1}
\eea
where $z= v \tau + ix$, $\tau$ being the imaginary time,
and $v \sim J_{\parallel} a_0$ denotes
the spin velocity on each chain ($a_0$ being the lattice spacing).

The next step to achieve the continuum limit of 
the Hamiltonian (\ref{hamiltmod1}) is to use the continuum 
representation of the spin densities of each chain\cite{affleck}:
\be 
\frac{{\bf S}_{a,j}}{a_0} \rightarrow
{\bf S}_a\left(x\right) = {\bf J}_a\left(x\right) +
 \left(-1\right)^{x/a_0}{\bf n}_a\left(x\right), \; a=0,1,..,N
\label{spinrep}
\ee
with $x=ja_0$ and 
${\bf J}_a$, ${\bf n}_a$ denote respectively the uniform and
staggered part of the spin density.
The smooth part ${\bf J}_a$ can be expressed as:
\be
{\bf J}_a = {\bf J}_{aR}  + {\bf J}_{aL},
\ee
whereas the ${\bf n}_a$ field is the vector part of the primary
field of the SU(2)$_1$ WZNW model transforming according to the 
fundamental representation of SU(2). 
Assuming weak interchain couplings $J_{\perp},
J_{\times} \ll J_{\parallel}$, one can perform
the continuum limit of the cylinder model using Eq. (\ref{hcin0})
and the continuum description (\ref{spinrep}) to obtain:
\be
{\cal H} = \frac{2\pi v}{3} \sum_{a=0}^{N}
\left({\bf J}_{aR} \cdot {\bf J}_{aR} +
{\bf J}_{aL} \cdot {\bf J}_{aL} \right) +
\gamma \sum_{a=0}^{N}{\bf J}_{aR} \cdot {\bf J}_{aL}
+ {\tilde g} {\bf n}_{0} \cdot {\bf n}_+
+  g {\bf J}_{0} \cdot {\bf I},
\label{conintmod1}
\ee
where we have discarded 
all irrelevant operators and oscillatory terms.
In Eq. (\ref{conintmod1}),
${\bf I}$ (respectively ${\bf n}_+$) is the sum 
of all uniform (respectively staggered) magnetizations 
of the surface chains:
\bea
{\bf I} &=& \sum_{a=1}^{N} {\bf J}_a \nonumber \\
{\bf n}_+ &=& \sum_{a=1}^{N} {\bf n}_a .
\label{sumcurrstag}
\eea
Moreover, the bare coupling constants in (\ref{conintmod1})
are given by: 
\bea 
g &=& a_0\left(J_{\perp} + 2 J_{\times}\right) \nonumber \\
{\tilde g} &=& a_0\left(J_{\perp} - 2 J_{\times}\right).
\label{barecoupmod1}
\eea

The interacting terms in (\ref{conintmod1})
are of different nature. A first one with the coupling 
constant ${\tilde g}$ is 
a relevant perturbation of scaling dimension $d=1$
and corresponds to the usual interchain interaction of 
non-frustrated spin ladders.
A second one with the coupling $g$ describes 
an interaction between the total spin current 
of the surface chains and that of the central chain.
This interaction is marginal relevant and, as long as ${\tilde g}$
is not too small, can be discarded. As a result, for generic 
values of ${\tilde g}$ and $g$, the low-energy 
physics of the model will be essentially governed by 
the backscattering contribution. 
In that case, frustration will play no 
role except for renormalization of spin velocities and mass gaps 
of the modes that will become massive in the IR limit. 
However, it is important to stress that, in constrast with
non-frustrated spin ladders, the two coupling 
constants ${\tilde g},g$ can vary independently in the model,
and there exists a vicinity of the line 
$J_{\perp} = 2 J_{\times}$ (${\tilde g}=0$) where frustration 
shows up in a nontrivial way.
Along this line, the low-energy properties of the model 
are mainly determined by current-current
interactions: 
\be
{\cal H} = \frac{2\pi v}{3} \sum_{a=0}^{N}
\left({\bf J}_{aR} \cdot {\bf J}_{aR} +
{\bf J}_{aL} \cdot {\bf J}_{aL} \right) +
\gamma \sum_{a=0}^{N}{\bf J}_{aR} \cdot {\bf J}_{aL}
+  g {\bf J}_{0} \cdot {\bf I},
\label{conintmod1bis}
\ee
which displays in the IR limit 
CSL physics as it will be shown in Section III.

\subsection{The (N+1)-chain cylinder model with doped central chain}

As already stated in the Introduction, the effect of doping might be 
useful to suppress the backscattering perturbation in the long 
distance limit. In particular, one can, for example, consider the previous 
(N+1)-chain cylinder model where some concentration of holes are introduced
in the central chain. Clearly, the very assumption that there can 
exist such model with only one chain doped needs justification.
Probably, the simplest one would be to assume that the 
intrasite single-electron energies on the central ($\epsilon_0$)
and surface ($\epsilon_S$) chains are much larger than the interchain
hopping amplitude: $t_{\perp} \ll |\epsilon_0 - \epsilon_S|$
so that different chains are out of resonance. Anyway, we shall 
assume in the following that one can dope the central chain only.
In that case, this special model belongs to the class of one-dimensional
electron liquid in an antiferromagnetic 
environment\cite{emery1,neto,granath,lehur} 
made by the surface chains. 

Due to the presence of doping, the continuum description of the spin 
density of the central is no longer given by Eq. (\ref{spinrep}) but takes now
the following form (see for instance chapter 16 of Ref. \cite{book1}):
\bea 
{\bf S}_0\left(x\right) = {\bf J}_0\left(x\right) +
 \cos\left(2k_{F,0} x + \sqrt{2\pi} \Phi_c\left(x\right)\right) 
{\bf n}_0\left(x\right), 
\label{spinrepcentrdop}
\eea
where $\Phi_c$ is a bosonic field associated with the charge degrees of 
freedom, $k_{F,0}$ being the Fermi momentum of the central chain.
In the continuum limit, the backscattering contribution (${\bf n}_0 
\cdot {\bf n}_+$) acquires now an oscillating 
factor $\sim \exp(i(2k_{F,0} - \pi)x)$
and thus becomes suppressed in the long distance limit if the concentration
of holes is sufficiently high.
As a result, the Hamiltonian of this asymmetric doped model in the 
spin sector is 
given by Eq. (\ref{conintmod1bis}) with only current-current interactions.

\subsection{The multichannel Kondo-Heisenberg model away from half-filling}

The last example that we shall give in this paper is 
the so-called Kondo-Heisenberg model with N-channel.
This model consists of N identical Hubbard chains (${\cal H}_U$)
interacting via a Kondo coupling with a periodic array of 
localized spins described by the spin-1/2 
operator ${\bf S}_{0,i}$\cite{white,sikkema,emery2,zachartse}.
The Hamiltonian of this system reads as follows:
\be
{\cal H} = {\cal H}_U + J_K \sum_i {\bf S}_{0,i} \cdot {\bf S}_{c,i}
+ J_H \sum_i {\bf S}_{0,i} \cdot {\bf S}_{0,i+1}
\label{Nchannel}
\ee
with: 
\be
{\cal H}_U  = -t \sum_{n,\sigma,i} \left(c^{\dagger}_{n\sigma,i}
c_{n\sigma,i+1} + h.c.\right) + U \sum_{n,i} c^{\dagger}_{n\uparrow,i}
c_{n\uparrow,i} c^{\dagger}_{n\downarrow,i} c_{n\downarrow,i},
\ee
where $c_{n\sigma,i}$  denotes the electron annihilation operator on
the site $i$, and $\sigma = \uparrow,\downarrow$, $n =1,..,N$
are respectively the electron spin and channel indices. The
conduction electron spin operator ${\bf S}_{c,i}$ at the ith site is defined
by:
\be
{\bf S}^{a}_{c,i} = \frac{1}{2} \sum_{n,\alpha,\beta}
c^{\dagger}_{n\alpha,i} {\sigma}^{a}_{\alpha\beta}
c_{n\beta,i}, \; a=x,y,z.
\ee
In Eq. (\ref{Nchannel}), the interaction between the electronic 
degrees of freedom and the localized spin is made by a Kondo 
coupling with the constant $J_K >0$. 
The Hamiltonian (\ref{Nchannel}) can also be viewed as a generalization
of the multichannel Kondo lattice with an exchange interaction $J_H >0$
between the impurities spins (simple RKKY interaction).

In the weak coupling limit when $J_K,U \ll t, J_H$, this Hamiltonian
can be represented in a bosonized form following the same route 
as in Refs. \cite{white,sikkema} for $N=1$.
The continuum description of the 
localized spin operator (${\bf S}_{0,i}$) is still given 
by Eq. (\ref{spinrep}) 
whereas that
of the conduction spin
density operator can be expressed in terms of N bosonic fields
$\Phi_{ac}$ and 
N SU(2)$_1$ spin currents ${\bf J}_{aR,L}$:
\bea
{\bf S}_c\left(x\right) = \sum_{a=1}^{N} \left({\bf J}_{aR}\left(x\right)
+ {\bf J}_{aL}\left(x\right) 
+\cos\left(2 k_F x + \sqrt{2\pi} \Phi_{ac}\left(x\right)\right)
{\bf n}_a\left(x\right)\right).
\label{condspin}
\eea

With these relations at hand, one 
can derive the continuum limit of the Hamiltonian (\ref{Nchannel})
away from half-filling (incommensurate filling). 
In that case, the $2k_F$ oscillation of the spin 
conduction operator becomes incommensurate 
with the alternating localized spin operator so that far away from 
half-filling, only marginal current-current interaction survives in
the Kondo interaction in the long distance limit:
\bea
{\cal H} = \frac{2\pi v_s}{3} \sum_{a=1}^{N} \left(
{\bf J}_{aR} \cdot {\bf J}_{aR} + {\bf J}_{aL} \cdot {\bf J}_{aL}\right)
+ \frac{2\pi v_0}{3} \left(
{\bf J}_{0R} \cdot {\bf J}_{0R} + {\bf J}_{0L} \cdot {\bf J}_{0L}\right)
\nonumber \\
+ g {\bf J}_0\cdot {\bf I} +
\gamma \sum_{a=1}^{N}{\bf J}_{aR} \cdot {\bf J}_{aL}
+ \gamma^{'} {\bf J}_{0R} \cdot {\bf J}_{0L}
\label{spinsector}
\eea
where ${\bf I} = \sum_{a=1}^{N} {\bf J}_a$, $v_s \sim 2t a_0 - 
Ua_0/2\pi$, $v_0 \sim J_H a_0$, $g\sim J_K a_0$, 
and $\gamma, \gamma^{'} <0$ ($U,J_H >0$).
Therefore, the
multichannel version of the Kondo-Heisenberg model
away from half-filling belongs to the class of models with marginally 
coupled current-current interactions. 
One should note that it is more transparent to use a different
basis, a spin-charge-flavor decomposition\cite{affleck0,affleck},
to single out the charge degrees of freedom.
In that case, the continuum description of the 
Hamiltonian of the N-channel Kondo-Heisenberg model
away from half filling reads as follows in the spin sector:
\bea
{\cal H}_s = \frac{2\pi v_s}{N+2}\left( 
{\bf I}_{R} \cdot {\bf I}_{R} + {\bf I}_{L} \cdot {\bf I}_{L}\right)
+ \frac{2\pi v_0}{3} \left(
{\bf J}_{0R} \cdot {\bf J}_{0R} + {\bf J}_{0L} \cdot {\bf J}_{0L}\right)
\nonumber \\
+ g {\bf J}_0\cdot {\bf I} +
\gamma {\bf I}_{R} \cdot {\bf I}_{L}
+ \gamma^{'} {\bf J}_{0R} \cdot {\bf J}_{0L}
\label{spinsectorbis}
\eea                                 

\section{Identification of the infrared fixed point}

We investigate now the IR properties of 
the Hamiltonian (\ref{conintmod1bis}) with only 
current-current interactions
which represents the continuum limit of several lattice models
as seen in the previous section. 
In particular, the nature of the IR fixed point that
governed the low-energy physics of these models for $N>1$ will be identified
and belongs to the universality class of chirally stabilized fluids 
introduced in Ref. \cite{andrei}.

Let us first look at the problem from a perturbative point of view
near the UV fixed point where all chains are decoupled so that
the model has the [SU(2)$_1$]$_R^{N+1}$ $\otimes$ [SU(2)$_1$]$_L^{N+1}$
symmetry. It is conformally invariant with a 
total central charge: $c_{UV}=N+1$ ($N+1$ gapless
bosonic modes). The one-loop Renormalization Group (RG) equations for 
the coupling constants $g$ and $\gamma$ are given by:
\bea 
\frac{d \gamma}{d \ln L} &=& \frac{\gamma^2}{2\pi v} \nonumber \\
\frac{d g}{d \ln L} &=& \frac{g^2}{2\pi v}.
\label{RGeq}
\eea
The bare coupling constant $\gamma$ is negative in all the models
presented in Section II, so 
that the perturbation $\gamma \sum_a {\bf J}_{aR} \cdot {\bf J}_{aL}$ 
is marginally irrelevant and can therefore be neglected: it
leads to logarithmic corrections.
On the other hand, the effective interchain coupling $g$ increases 
upon renormalization ($g >0$).
Usually, the development of a strong coupling regime ($g(L) \rightarrow 
+\infty$) is accompanied by a dynamical mass generation and the 
loss of conformal invariance as for the non-linear sigma model or 
the Gross-Neveu model in 1+1 dimensions.
This is indeed the case for $N=1$ which represents a continuum
version of the two-leg zigzag ladder\cite{white,allen} 
in the absence of the so-called twist
perturbation\cite{nersesyan1}.
However, for $N\ge 2$, the interaction 
flows to an intermediate fixed point ($g^{*} < \infty$) where
the system displays critical properties with a smaller (and
generally non-integer) central charge than $c_{UV}$
according to the Zamolodchikov c-theorem\cite{zamolo1}.

To this end, let us rewrite the Hamiltonian (\ref{conintmod1bis})
in terms of the leading part which governs 
the strong coupling behaviour:
\be
{\cal H} = \frac{2\pi v_1}{3} \sum_{a=1}^{N}
\left({\bf J}_{aR} \cdot {\bf J}_{aR} +
{\bf J}_{aL} \cdot {\bf J}_{aL} \right) +
\frac{2\pi v_0}{3}
\left({\bf J}_{0R} \cdot {\bf J}_{0R} +
{\bf J}_{0L} \cdot {\bf J}_{0L} \right)
+ g \left({\bf J}_{0R} \cdot {\bf I}_L + {\bf J}_{0L} \cdot {\bf I}_R\right)
\label{hgenstrong}
\ee
where we have omitted the marginal irrelevant contribution and the 
interaction between the currents of the same chirality whose effect
will be effectively taken into account by 
allowing the surface chains velocity ($v_1$)
to be different from that of the central chain ($v_0$).
We shall 
explicitely check for $N=2$, in Appendix D, by our Toulouse point approach 
that all these discarded interactions will not destabilize the intermediate
fixed point.

The structure (\ref{hgenstrong}) for $N>1$ suggests
that some degrees of freedom do not participate to the interaction 
and thus will decouple and remain massless.
Indeed, the free part of the Hamiltonian (\ref{hgenstrong}) is a sum of $(N+1)$ 
critical SU(2)$_1$ WZNW models whereas the interaction describes
a coupling between the SU(2)$_1$ spin current of the central 
chain (${\bf J}_{0R,L}$) 
and the total surface current (${\bf I}_{R,L}$)
which is a SU(2)$_N$ current being the sum of $N$ SU(2)$_1$ currents
(see Eq. (\ref{sumcurrstag})).
To identify the nature of the non-interacting degrees of freedom,
we use the following decomposition:
\be
\prod_{i=1}^{N} SU(2)_1|_i = SU(2)_N \otimes {\cal G}_N
\label{cosetN}
\ee
where ${\cal G}_N$ is some piece.
The central charge of the critical model with symmetry ${\cal G}_N$
is: 
\be 
c_{{\cal G}_N} = N - c_{SU(2)_N} = 
N - \frac{3N}{N + 2}=
\frac{N \left(N - 1\right)}{N + 2}
\label{centdiscr}
\ee
where we have used that the central charge of the SU(2)$_k$ WZNW
model is: $c=3k/(k+2)$\cite{knizhnik}.
Though it requires a proof, for $N=2m$, ${\cal G}_N$ 
can be viewed as a direct 
product of $m$ $Z_{2m}$-symmetric critical models\cite{zamolo2}
whereas for $N = 2 m + 1$, it is the product of $m$ $Z_{2m}$-symmetric
critical models and the minimal model ${\cal M}_{N+1}$.
We recall here that the central charge of the minimal model series
${\cal M}_{p}$ with $p\ge 2$ is: $c_p = 1- 6/p(p+1)$\cite{polyakov,dms}. 
The previous identification reproduces the value of the central charge
and also all primary fields have the correct scaling dimensions
but it is not sufficient to prove the previous CFT embedding.
Anyway, in this work, we only need the actual value of the 
central charge (\ref{centdiscr}) of the critical model
with symmetry ${\cal G}_N$ which one can note that it coincides with
the sum of the central charge of the $N-1$ first minimal models:
\be
c_{{\cal G}_N} = \frac{N\left(N-1\right)}{N+2}
= \sum_{m=2}^{N+1}\left(1 - \frac{6}{m\left(m+1\right)}\right).
\ee

The actual model under consideration simplifies thus in the
following way: 
\be 
{\cal H} = {\cal H}_{{\cal G}_N} +  \bar{\cal H}
\label{HsimN}
\ee
where ${\cal H}_{{\cal G}_N}$ is the Hamiltonian corresponding
to the critical degrees of freedom with symmetry ${\cal G}_N$
that decouple from the interaction.
All non-trivial physics is incorporated in
the current dependent part of the Hamiltonian ($\bar{\cal H}$)
which can be decomposed into a sum of two commuting
and {\it chirally asymmetric} parts:
\be
\bar{\cal H} = {\cal H}_1 + {\cal H}_2, ~([{\cal H}_1, {\cal H}_2] = 0)
\label{hambarN0}
\ee
where
\bea
{\cal H}_1 &=&  \frac{2\pi v_1}{N+2} \;  {\bf I}_{R} \cdot{\bf I}_{R} +
 \frac{2\pi v_0}{3} \; {\bf J}_{0L} \cdot{\bf J}_{0L}
+ g \;  {\bf I} _{R} \cdot {\bf J}_{0L} \nonumber \\
\nonumber \\
{\cal H}_2 &=&  \frac{2\pi v_1}{N+2} \;  {\bf I}_{L} \cdot{\bf I}_{L} +
 \frac{2\pi v_0}{3} \; {\bf J}_{0R} \cdot{\bf J}_{0R}
+ g \;  {\bf I} _{L} \cdot {\bf J}_{0R}.
\label{hamN1}
\eea
The Hamiltonian ${\cal H}_1$ thus describes two marginally
coupled [SU(2)$_N$]$_R$ and [SU(2)$_1$]$_L$ WZNW models
and ${\cal H}_2$ is obtained from ${\cal H}_1$ by interchanging
the indices $R$ and $L$.

The properties of chirally asymmetric models with the structure of
${\cal H}_1$ in Eq. (\ref{hamN1}) 
but with a 
more general symmetry group [SU(2)$_{f_{R}}$]$_R$ 
$\otimes$ [SU(2)$_{f_{L}}$]$_L$
($f_R > f_L$) have been studied
by Andrei et al.\cite{andrei} (see Eq. (\ref{handrei})).
These authors argued that
such models flow in the IR limit to an intermediate fixed point
$g^{*}$ 
(chiral fixed point) whose symmetry is determined by a
WZNW coset\cite{gko}:
\be
\left[SU(2)_{f_R - f_L}\right]_{R} \otimes
\left[\frac{SU(2)_{f_L}\times
SU(2)_{f_R - f_L}}{SU(2)_{f_R}}\right]_{L}.
\label{coset}
\ee
This fixed point has been found by the authors of
Ref. \cite{andrei} using the fact that
the differences between the chiral
components of the central charge ($c_R-c_L$)
and the KM levels ($f_R-f_L$) are preserved under the RG flow.
Morevover, since the total central charge ($c_R+c_L$)
always decreases upon renormalization according to the
Zamolodchikov c-theorem\cite{zamolo1}, the identification of the IR fixed point
is equivalent to find a conformally invariant model
with the lowest total central charge consistent with
the fixed values of $c_R-c_L$ and $f_R-f_L$.
It turns out that the solution for this problem
is given by the coset model (\ref{coset}).
The structure of this fixed point has also been checked 
by the computation of the low temperature behaviour of the specific heat 
from the Thermodynamical Bethe ansatz approach\cite{andrei} since 
this physical quantity is a direct probe of the value of the 
central charge of the IR fixed point\cite{cardy}.

Using the result (\ref{coset}), one can 
immediately deduce the structure of the IR
fixed point of the Hamiltonian (\ref{hambarN0}):
\bea
\left[\left[SU(2)_{N - 1}\right]_{R} \otimes
\left[\frac{SU(2)_{1}\times
SU(2)_{N - 1}}{SU(2)_{N}}\right]_{L}\right]
\otimes 
\left[\left[SU(2)_{N - 1}\right]_{L} \otimes
\left[\frac{SU(2)_{1}\times
SU(2)_{N - 1}}{SU(2)_{N}}\right]_{R}\right] \nonumber\\
\label{groupsymmhbar}
\eea
where the first (respectively second) part is the symmetry of the 
IR fixed point associated with ${\cal H}_1$ (respectively 
${\cal H}_2$).
The central charge in the IR limit (${\tilde c}_{IR}$) 
corresponding to the model
described by the 
Hamiltonian ($\bar{\cal H}$) of Eq. (\ref{hambarN0}) is thus given
by:
\be 
{\tilde c}_{IR}\left(N\right) 
= \frac{2\left(N-1\right)\left(2N+5\right)}{\left(N+1\right)
\left(N+2\right)}.
\label{centralcharghbar}
\ee
From this result, we deduce the value of the
total central at the IR fixed point of 
the original Hamiltonian (\ref{hgenstrong}) taking into account the contribution
of the massless degrees of freedom (\ref{centdiscr}) that do not participate 
to the interaction:
\be
c_{IR}\left(N\right) = {\tilde c}_{IR}\left(N\right) + c_{{\cal G}_N} = 
\frac{\left(N-1\right)\left(N^2+5N + 10\right)}{\left(N+1\right)
\left(N+2\right)}. 
\label{totalcentralchargeNfin}
\ee

In summary, the low-energy physics of the (N+1)-chain cylinder model,
its asymmetric doped version, and 
the multichannel Kondo-Heisenberg model away from half-filling,
are governed by an intermediate
IR fixed point displaying CSL behaviour with a central charge 
which is not integer but rational for a generic value of $N$.
The first two models, in the spin sector, have a 
central charge given by 
Eq. (\ref{totalcentralchargeNfin}) whereas the gapless modes 
corresponding to the {\sl spin} sector of the N-channel 
Kondo-Heisenberg model are fixed by Eq. (\ref{centralcharghbar}).
Moreover, from Eqs. (\ref{centralcharghbar},\ref{totalcentralchargeNfin}), 
one can notice that the
central charges vanishe for $N=1$ as it should be (spin gap phase)
whereas for $N=2$ one has the simple values:
${\tilde c}_{IR}(2)= 3/2, c_{IR}(2) = 2$ which means
that a simple and independent approach should be possible
in that case.

In the following, we shall focus on this $N=2$ special case by 
presenting an exact solution of the U(1) version of the $N=2$ 
problem using a Toulouse point solution.
This approach has been extremely fruitful especially in 
Quantum Impurity problems and gives a simple description of the 
two-channel Kondo model and also of the 
Kondo lattice\cite{emery,zachar,georges}.
In particular, for the two-channel Kondo problem, Emery 
and Kivelson\cite{emery} identified the residual zero point entropy 
stemming from the decoupling of a Majorana fermion degrees of freedom
using a mapping of the model onto free fermion theory for a very
special value of the interaction (Toulouse point).
Although the position of the solvable point is non-universal, the 
Toulouse limit solution captures the physical and universal properties
of the low-temperature behaviour of the model.
In our problem, the Toulouse point approach provides a {\sl direct and 
transparent} reading of the spectrum of the model for $N=2$.
It also gives a 
non-perturbative basis from which all leading asymptotics of spin-spin
correlations for the 3-chain cylinder model can be determined. 

\section{The Toulouse point solution in the N=2 case}

In this section, we present the solution,
by a Toulouse point approach, 
of the model (\ref{hgenstrong}) with $N=2$:
\bea
{\cal H} = \frac{2 \pi v_1}{3} \;\left({\bf J}_{1R}.{\bf J}_{1R}
+ {\bf J}_{1L}.{\bf J}_{1L}+
{\bf J}_{2R}.{\bf J}_{2R} + {\bf J}_{2L}.{\bf J}_{2L} \right)
+ \frac{2 \pi v_0}{3} \;\left({\bf J}_{0R}.{\bf J}_{0R}
+ {\bf J}_{0L}.{\bf J}_{0L}\right)
\nonumber \\
+ \; g \;\left({\bf I}_R. {\bf J}_{0L}
+ {\bf I}_L. {\bf J}_{0R} \right).
\label{hsu2}
\eea

\subsection{Decoupling of a Z$_2$ non-magnetic excitation}

As discussed in section III A, 
the structure of the interaction of Eq. (\ref{hsu2}) suggests an alternative 
representation of the two SU(2)$_1$ WZNW models of the surface 
chains in a SU(2)$\otimes$ Z$_2$ way reflecting the global 
spin rotational symmetry as well as the discrete Z$_2$ 
related to the interchange symmetry ($1 \rightarrow 2$). 
The critical model with the symmetry 
group ${\cal G}_2$ in the description (\ref{cosetN}) for $N=2$ 
identifies with a single critical Ising model (or a free Majorana
fermion in the continuum limit) since on has from  
the Goddard-Kent-Olive
construction\cite{gko}: 
\be
SU(2)_1 \otimes SU(2)_1 \sim SU(2)_2 \otimes Z_2.  
\label{gkoe}
\ee
A simple way to take into account this non-magnetic Z$_2$ 
excitation (Ising degrees of freedom)
is to use the representation of two SU(2)$_1$ spin currents 
in terms of four Majorana fermions 
$\xi^{0}$ and $\vec \xi$\cite{zamolo,allen,book1}: 
\bea
{\bf I}_{\alpha} = 
{\bf J}_{1\alpha} + {\bf J}_{2\alpha}
= -\frac{\ri}{2} \; 
{\vec \xi}_{\alpha} \wedge {\vec \xi}_{\alpha}, \; \alpha=R,L, \nonumber\\
{\bf J}_{1\alpha} - {\bf J}_{2\alpha} = \ri
\; {\vec \xi}_{\alpha} \xi_{\alpha}^{0}, \; \alpha=R,L
\label{j1mj2}
\eea
where the four right and left moving Majorana fermions have 
the following anticommutation relations ($a,b=0,1,2,3$):
\bea
\{\xi_{R,L}^a(x),\xi_{R,L}^b(y)\} &=& \delta^{ab}\delta(x - y) \nonumber \\
\{\xi_R^a(x),\xi_L^b(y)\} &=& 0
\eea
and are normalized according to:
\bea 
\xi^a_L\left(z\right) \xi^b_L\left(\omega\right)
&\sim& \frac{\delta^{ab}}{2\pi\left(z-\omega\right)} \nonumber \\
\xi^a_R\left(\bar z\right) \xi^b_R\left(\bar \omega\right)
&\sim& \frac{\delta^{ab}}{2\pi\left(\bar z-\bar \omega\right)} \nonumber \\
\xi^a_R\left(\bar z\right) \xi^b_L\left(\omega\right)
&\sim& 0 . 
\eea
Using the correspondence (\ref{j1mj2}), 
the Hamiltonian (\ref{hsu2}) simplifies as 
\be
{\cal H} = - \ri \frac{v_1}{2} \left(\xi_{R}^{0} \partial_x  \xi_{R}^{0} -
\xi_{L}^{0} \partial_x  \xi_{L}^{0}  \right)
+ \bar{\cal H} [{\bf I}, {\bf J}_0]
\label{htotexact}
\ee
where the first part is the free Hamiltonian of the Majorana 
fermion $\xi^0$ with central charge $c=1/2$ and
\be 
\bar{\cal H} = 
\frac{\pi v_1}{2} \;\left({\bf I}_{R}.{\bf I}_{R} + 
{\bf I}_{L}.{\bf I}_{L}\right) + 
\frac{2 \pi v_0}{3} \;\left({\bf J}_{0R}.{\bf J}_{0R}
+ {\bf J}_{0L}.{\bf J}_{0L}\right)
+ \; g \;\left({\bf I}_R. {\bf J}_{0L}  
+ {\bf I}_L. {\bf J}_{0R} \right). 
\label{hbarsu2}
\ee
All non-trivial physics is incorporated in 
the current dependent part of the Hamiltonian ($\bar{\cal H}$) 
which decomposes into a sum of two commuting 
and chirally asymmetric parts as in 
Eqs. ((\ref{hambarN0}), (\ref{hamN1})) for $N=2$: 
\be
\bar{\cal H} = {\cal H}_1 + {\cal H}_2, ~([{\cal H}_1, {\cal H}_2] = 0)
\label{hambar0}
\ee
\bea
{\cal H}_1 &=&  \frac{\pi v_1}{2} \;  {\bf I}_{R} \cdot{\bf I}_{R} +
 \frac{2\pi v_0}{3} \; {\bf J}_{0L} \cdot{\bf J}_{0L}
+ g \;  {\bf I} _{R} \cdot {\bf J}_{0L} \nonumber \\
{\cal H}_2 &=&  \frac{\pi v_1}{2} \;  {\bf I}_{L} \cdot{\bf I}_{L} +
 \frac{2\pi v_0}{3} \; {\bf J}_{0R} \cdot{\bf J}_{0R}
+ g \;  {\bf I} _{L} \cdot {\bf J}_{0R}. 
\label{ham1}
\eea

\subsection{Identification of the Toulouse point}

We consider now 
the SU(2) broken version of model (\ref{hambar0}) ($g 
\rightarrow g_{\parallel}, g_{\perp}$) where 
we allow $g_{\parallel}$ to be different from 
$g_{\perp}$: 
\be
{\cal H}_1 = \frac{\pi v_1}{2} \;{\bf I}_R.{\bf I}_R
+ \frac{2 \pi v_0}{3} \;{\bf J}_{0L}.{\bf J}_{0L}
+ \; g_{\parallel} \;I_R^z J_{0L}^z
+ \frac{g_{\perp}}{2} \;\left(I_R^{+} J_{0L}^{-}
+ H.c. \right)
\label{hu1}
\ee
and ${\cal H}_2$ is obtained from ${\cal H}_1$ by
inverting chiralities of all the spin currents.

In this section, we present an exact solution of 
the Hamiltonian (\ref{hu1}) using 
Abelian bosonization. 
The solution is based on a mapping onto Majorana fermions 
and exploits the existence of a Toulouse-like, exactly 
solvable point,  
at a special value (though non-universal) of $g_{\parallel}$ 
where the fermions are free. 

The scaling equations for the model (\ref{hu1}) 
are of the Kosterlitz-Thouless
form:
\be
\frac{d g_{\parallel}}{d \ln L} =
\frac{g_{\perp}^2}{\pi\left(v_1+v_0\right)},
\; \frac{d g_{\perp}}{d \ln L} =
\frac{g_{\parallel}g_{\perp}}{\pi\left(v_1+v_0\right)},
\label{rgxxz}
\ee
indicating the increase of the coupling constants upon
renormalization for $g_{\parallel} > - |g_{\perp}|$. Since
both the exactly solvable point and the isotropic strong-coupling
separatrix $g_{\parallel} = |g_{\perp}|$ occur within this range,
the exact solution of the anisotropic model (\ref{hu1}) is expected
to exhibit generic properties of the original, SU(2) symmetric,
model (\ref{ham1}).                                            

The starting point of our approach is the Abelian bosonization 
of the component of the SU(2)$_1$ current ${\bf J}_0$
associated with the spin excitation of the central chain.
We introduce a massless bosonic field ${\varphi}$ 
and write (see for instance Appendix A of Ref. \cite{shelton}
or chapter 13 of Ref. \cite{book1}):
\bea 
J_{0R}^z &=& \frac{1}{\sqrt{2\pi}}\; \partial_x \varphi_R, \;\; \;
\; \;\;\; \;
J_{0L}^z = \frac{1}{\sqrt{2\pi}}\; \partial_x \varphi_L \nonumber \\ 
J_{0R}^{\pm} &=& \frac{1}{2\pi a_0}\; \re^{\mp \ri \sqrt{8\pi} \varphi_R}, 
\;\; J_{0L}^{\pm} = \frac{1}{2\pi a_0}\; \re^{\pm \ri \sqrt{8\pi} \varphi_L}.
\label{boso1}
\eea 
It is a simple matter to show that the representation (\ref{boso1}) indeed 
reproduces 
the OPEs (\ref{opesu2level1}) of the SU(2)$_1$ WZNW model

On the other hand, the SU(2)$_2$ current ${\bf I}$ can be 
expressed in terms of a triplet of massless Majorana fermions $\xi^a$,
a=1,2,3 (see Eq. (\ref{j1mj2})).  
Since we consider the U(1)$\otimes Z_2$-symmetric version of the 
model, it is natural to discriminate $(\xi^1, \xi^2)$ from 
the $\xi^3$ field. The two Majorana fields $(\xi^1, \xi^2)$ can be 
combined to form a single Dirac field $\chi$ which can in turn 
be bosonized with the introduction of a massless bosonic 
field $\Phi$: 
\bea
\chi_R &=& \frac{\xi_R^2 + \ri \xi_R^1}{\sqrt 2} = 
\frac{\kappa}{\sqrt{2\pi a_0}} \;
\re^{\ri \sqrt{4\pi} \Phi_R} \nonumber \\
\chi_L &=& \frac{\xi_L^2 + \ri \xi_L^1}{\sqrt 2} = 
\frac{\kappa}{\sqrt{2\pi a_0}} \;
\re^{-\ri \sqrt{4\pi} \Phi_L}, 
\label{boso2}  
\eea
where the anticommutation relation 
between $\chi_R$ and $\chi_L$ is insured by the 
choice $[\Phi_R, \Phi_L]=\ri/4$. On the other hand, to take into account
the correct
anticommutation with the third Majorana fermion $\xi^3$, one 
needs to introduce 
an additional real fermionic degree 
of freedom $\kappa$ ($\kappa^2=1$). Using Eqs. (\ref{j1mj2}, \ref{boso2}),
the components of the SU(2)$_2$ spin current ${\bf I}$ transform to:
\bea 
I_R^z &=& \frac{1}{\sqrt{\pi}}\; \partial_x \Phi_R, \;\;\;\;\;\; 
\;\;\;\;\;\;\;\;\;\;
I_L^z = \frac{1}{\sqrt{\pi}}\; \partial_x \Phi_L \nonumber \\ 
I_R^{\pm} &=& \frac{\ri}{\sqrt{\pi a_0}}\;\xi_R^3 \kappa
\re^{\mp \ri \sqrt{4\pi} \Phi_R}, \; 
I_L^{\pm} = \frac{\ri}{\sqrt{\pi a_0}}\;\xi_L^3 \kappa
\re^{\pm \ri \sqrt{4\pi} \Phi_L}. 
\label{boso3}
\eea 
The chiral aymmetric parts 
of the total 
Hamiltonian $\bar{\cal H}$ (\ref{hambar0}) can then be written 
in the following
bosonized form: 
\bea
{\cal H}_1 = v_0 \; \left(\partial_x \varphi_L\right)^2 + 
v_1 \; \left(\partial_x \Phi_R\right)^2 -\ri \frac{v_1}{2} \;
\xi_R^3 \partial_x \xi_R^3 \nonumber \\
+ \frac{g_{\parallel}}{\sqrt 2 \pi} \; \partial_x \varphi_L 
\partial_x \Phi_R + \frac{\ri g_{\perp}}{2 \left(\pi a_0\right)^{3/2}} 
\; \xi_R^3 \kappa \cos \left(\sqrt{4\pi} \Phi_R + \sqrt{8\pi} \varphi_L
\right)
\label{h1boso}
\eea
and 
\bea
{\cal H}_2 = v_0 \; \left(\partial_x \varphi_R\right)^2 +
v_1 \; \left(\partial_x \Phi_L\right)^2 +\ri \frac{v_1}{2} \;
\xi_L^3 \partial_x \xi_L^3 \nonumber \\
+ \frac{g_{\parallel}}{\sqrt 2 \pi} \; \partial_x \varphi_R
\partial_x \Phi_L + \frac{\ri g_{\perp}}{2 \left(\pi a_0\right)^{3/2}} 
\; \xi_L^3 \kappa \cos \left(\sqrt{4\pi} \Phi_L + \sqrt{8\pi} \varphi_R
\right).
\label{h2boso}
\eea

One can eliminate the cross terms $\partial_x \varphi \partial_x \Phi$ 
by performing a canonical transformation: 
\bea 
\left(\begin{array}{c}
\varphi_L\\
\Phi_R
\end{array}\right) = 
\left(
\begin{array}{lccr}
{\rm ch} \alpha & {\rm sh} \alpha \\
{\rm sh} \alpha & {\rm ch} \alpha
\end{array} \right) \left(\begin{array}{c}
\bar \Phi_{2L}\\
\bar \Phi_{1R}
\end{array}\right) 
\label{can1}
\eea
and
\bea
\left(\begin{array}{c}
\Phi_L\\
\varphi_R
\end{array}\right) =
\left(
\begin{array}{lccr}
{\rm ch} \alpha & {\rm sh} \alpha \\
{\rm sh} \alpha & {\rm ch} \alpha
\end{array} \right) \left(\begin{array}{c}
\bar \Phi_{1L}\\
\bar \Phi_{2R}
\end{array}\right)
\label{can2}
\eea
with 
\be 
{\rm th} 2 \alpha = - \frac{g_{\parallel}}{\pi \sqrt 2 \left(v_1+v_2\right)}.
\label{th1}
\ee
Under the transformation (\ref{can1}) and (\ref{can2}), the 
arguments of the cosines in Eqs. (\ref{h1boso},\ref{h2boso}) 
become: 
\bea 
\sqrt{4\pi} \Phi_R + \sqrt{8\pi} \varphi_L
\rightarrow \sqrt{4\pi} \left[\left(
\sqrt 2 {\rm ch} \alpha + {\rm sh} \alpha\right) \bar \Phi_{2L} 
+ \left({\rm ch} \alpha + \sqrt 2 {\rm sh} \alpha \right) \bar \Phi_{1R} 
\right] \\ 
\sqrt{4\pi} \Phi_L + \sqrt{8\pi} \varphi_R
\rightarrow \sqrt{4\pi} \left[\left(
\sqrt 2 {\rm ch} \alpha + {\rm sh} \alpha\right) \bar \Phi_{2R} 
+ \left({\rm ch} \alpha + \sqrt 2 {\rm sh} \alpha \right) \bar \Phi_{1L} 
\right].
\label{costran}
\eea
We immediately observe that for a special
value of $\alpha$ given by 
\be
{\rm th} \alpha = -\frac{1}{\sqrt 2},
\label{th2}
\ee
we are at the free fermion point where 
the two cosine terms acquire respectively conformal weights $(0,1/2)$
and $(1/2,0)$ and 
can be expressed in terms of the left and right components of 
a single Dirac fermion $\psi$. When $\alpha$ 
satisfies both Eqs. (\ref{th1}, \ref{th2}), $g_{\parallel}$ takes 
a special $positive$ value $g_{\parallel}^{*}$ (the Toulouse point): 
\be
g_{\parallel}^{*} = \frac{4\pi}{3} \left(v_1+v_0\right), 
\label{toulouse}
\ee
and the total Hamiltonian ($\bar{\cal H}$) reads as follows:
\bea 
\bar {\cal H} &=& - \frac{\ri v_1}{2}\left(\xi_R^{3}\partial_x \xi_R^{3} 
-\xi_L^{3}\partial_x \xi_L^{3}\right) + 
\frac{u_1}{2} \left[\left(\partial_x {\bar \Phi}_1\right)^2 + 
\left(\partial_x  {\bar \Theta}_1\right)^2 \right] +
\frac{u_2}{2} \left[\left(\partial_x {\bar \Phi}_2\right)^2 + 
\left(\partial_x  {\bar \Theta}_2\right)^2 \right] \nonumber \\
&+& \frac{\ri g_{\perp}}{2 \left(\pi a_0\right)^{3/2}}
\; \xi_R^3 \kappa \cos \left(\sqrt{4\pi} \bar \Phi_{2L}\right) 
+ \frac{\ri g_{\perp}}{2 \left(\pi a_0\right)^{3/2}}
\; \xi_L^3 \kappa \cos \left(\sqrt{4\pi} \bar \Phi_{2R}\right) 
\label{htrans}
\eea
where we have reestablished the complete bosonic
fields $\bar \Phi_{a}= \bar \Phi_{aL} + \bar \Phi_{aR}$ ($a=1,2$)
and their dual counterpart $\bar \Theta_{a} = 
\bar \Phi_{aL} - \bar \Phi_{aR}$. 
The associated velocities $u_1$ and $u_2$ 
are given at the Toulouse point by:
\be
u_1 = \frac{2 v_1 - v_0}{3},\ \  u_2 = \frac{2 v_0 - v_1}{3}.
\label{velo}
\ee
The Toulouse point solution is thus stable provided 
that all the velocities of the modes are positive, i.e. 
$1/2 \le v_0/v_1 \le 2$.
Since we are at the free fermion point, we can 
refermionize the two cosine terms in Eq. (\ref{htrans}). 
Introducing 
a pair of Majorana fields $\eta$ and $\zeta$ and using 
the correspondence:
\bea
\psi_{R,L} &=& \frac{\eta_{R,L}+ i \zeta_{R,L}}{\sqrt 2} \nonumber\\
           &=& \frac{\kappa}{\sqrt{2\pi a_0}} \;
\re^{\pm \ri \sqrt{4\pi} \bar \Phi_{2R,L}},
\label{psi}
\eea
the two cosine terms become simply:
\be
\frac{\kappa}{\sqrt{\pi a_0}} \cos \left(\sqrt{4\pi} 
\bar \Phi_{2R,L}\right) = 
\eta_{R,L}.
\ee
Therefore, the Hamiltonian ($\bar{\cal H}$) takes the following final form:
\bea
{\bar {\cal H}} &=& \frac{u_1}{2} \left[  (\partial_x {\bar \Phi}_1)^2 +
 (\partial_x  {\bar \Theta}_1)^2 \right] - \frac{\ri u_2}{2}
 \left[ \zeta_R\partial_x \zeta_R - \zeta_L\partial_x \zeta_L\right]
 - \frac{\ri v_1}{2}\left[ \xi_R^{3}\partial_x \xi_R^{3}
-\xi_L^{3}\partial_x \xi_L^{3}\right]\nonumber\\
&-& \frac{\ri u_2}{2}
 \left[ \eta_R\partial_x \eta_R - \eta_L\partial_x \eta_L\right] +
 \ri m \left[ \xi_R^{3}\eta_L - \eta_R\xi_L^{3}\right]
\label{hfin}
\eea
with $m = g_{\perp}/2 a_0 \pi$. 
Taking into account the contribution of the $Z_2$
Majorana fermion $\xi^0$ which
has decoupled from the beginning,
the original Hamiltonian (\ref{htotexact}) reads at the Toulouse point:
\bea
{\cal H} &=& \frac{u_1}{2} \left[  (\partial_x {\bar \Phi}_1)^2 +
 (\partial_x  {\bar \Theta}_1)^2 \right] - \frac{\ri u_2}{2}
 \left[ \zeta_R\partial_x \zeta_R - \zeta_L\partial_x \zeta_L\right]
 - \frac{\ri v_1}{2}\left[ \xi_R^{0}\partial_x \xi_R^{0}
-\xi_L^{0}\partial_x \xi_L^{0}\right]\nonumber\\
 &-& \frac{\ri v_1}{2}\left[ \xi_R^{3}\partial_x \xi_R^{3}
-\xi_L^{3}\partial_x \xi_L^{3}\right]
- \frac{\ri u_2}{2}
 \left[ \eta_R\partial_x \eta_R - \eta_L\partial_x \eta_L\right] +
 \ri m \left[ \xi_R^{3}\eta_L - \eta_R\xi_L^{3}\right].
\label{hfintot}
\eea

At the Toulouse point, the massless degrees of freedom are thus 
described in terms of an effective S=1/2 Heisenberg spin chain 
associated with the ${\bar \Phi}_1$ field and two decoupled critical
Ising models ($\xi^0, \zeta$) so that the total central 
charge in the IR limit is: $c_{IR}=2$
in full agreement with Eq. (\ref{totalcentralchargeNfin}) for $N=2$. 
One should also notice that the nature of the critical fields found in
the Toulouse limit approach reproduces, for $N=2$, the
structure of the symmetry
group (\ref{groupsymmhbar}) of the fixed point
corresponding to the Hamiltonian
(${\bar {\cal H}}$).
Indeed, substituting the value $N = 2$ in Eq. (\ref{groupsymmhbar}),
we see
using Eq. (\ref{gkoe}) that
this symmetry group simplifies as:
\be
\left[\left[SU(2)_1\right]_R \otimes \left[Z_2 \right]_L \right]
\otimes
\left[\left[SU(2)_1\right]_L \otimes \left[Z_2 \right]_R \right]
\ee
which is in
full agreement with the
structure of the Hamiltonian (\ref{hfin}) at the Toulouse point.

The remaining part of the Hamiltonian (\ref{hfintot}) 
has a spectral gap and 
describe the hybridization of the Majorana $\xi^3$ and $\eta$ fields with
different chiralities. Indeed, it can be written as a sum of two commuting 
part ${\cal H}_{+}+ {\cal H}_{-}$ with
\bea
{\cal H}_{+} &=& \frac{\ri v_1}{2}\xi_L^{3}\partial_x \xi_L^{3}
 -\frac{\ri u_2}{2}
\eta_R\partial_x \eta_R
-\frac{\ri m}{2} \left(\eta_R\xi_L^{3} - \xi_L^{3} \eta_R\right) \nonumber \\
{\cal H}_{-} &=& -\frac{\ri v_1}{2}\xi_R^{3}\partial_x \xi_R^{3}
+\frac{\ri u_2}{2}
\eta_L\partial_x \eta_L
+\frac{\ri m}{2} \left(\xi_R^{3}\eta_L - \eta_L\xi_R^{3}\right).
\label{hybrid}
\eea
The resulting spectrum is given by:
\bea
E_{+}\left(k\right) &=& k u_1 \pm \sqrt{k^2\left(\frac{v_1+v_0}{3}\right)^2 +
m^2} \label{hybrispecp}\\
\nonumber\\
E_{-}\left(k\right) &=& -k u_1 \pm \sqrt{k^2\left(\frac{v_1+v_0}{3}\right)^2 +
m^2}, \label{hybrispecm}
\eea
which, apart from the velocity anisotropy, is reminiscent
of the structure of excitations found in the 
Majorana approach for the Kondo lattice\cite{coleman}. 
One should stress that each Majorana fermion hybridizes with the 
other with the opposite chirality and that there is no coupling between 
Majorana fermion of the same chirality. This reflects the chiral 
nature of the fixed point. 
One should finally note that our solution
contains the SU(2) point when $g_{\perp} = 
g_{\parallel}$ and the mass at the Toulouse 
point reads in that case: $m = g_{\parallel}^{*}/2\pi a_0 =
2(v_1+v_0)/3a_0$. 

The exact solution based on the existence 
of a Toulouse point allows us to get a simple understanding 
of the structure of the critical elementary excitations and 
to estimate the leading asymptotics of 
the correlation functions. 
In the following, we shall realize this program by directly focusing 
on the physical properties of the 3-chain cylinder model.  

\section{Physical properties of the 3-chain cylinder model} 

In this section, the spectral 
properties of the 3-chain cylinder model 
are discussed and  we
compute the leading asymptotics of all spin-spin
correlation functions at the chiral fixed point. 
To this end, we first express the smooth and staggered 
magnetization of each chain in terms of the suitable 
fields describing the long distance physics at the 
chiral fixed point. 

\subsection{Representation of the spin currents and 
staggered spin fields at the Toulouse point}

In this subsection, the spin currents (${\bf J}_{aR,L}$) 
and staggered spin fields (${\bf n}_a$) of the 
chain $a=0,1,2$ are related 
to the two critical Majorana fermions ($\xi^0, \zeta$) 
and the massless bosonic field ($\bar \Phi_1$) at the Toulouse 
point. To get a better insight in the structure of excitations,
we introduce the smooth part (${\cal J}^a$) and 
staggered part (${\cal N}^a$) corresponding to 
the effective S=1/2 Heisenberg spin chain described by
the bosonic $\bar \Phi_1$ field. Using Eq. (\ref{boso1}),
one has the following correspondence:
\bea
{\cal J}_R^z &=& \frac{1}{\sqrt{2\pi}}\; \partial_x \bar \Phi_{1R}, \;\; \;
\; \;\;\; \;
{\cal J}_L^z = \frac{1}{\sqrt{2\pi}}\; \partial_x \bar \Phi_{1L} \nonumber \\
{\cal J}_R^{\pm} &=& \frac{1}{2\pi a_0}\; \re^{\mp \ri \sqrt{8\pi} 
\bar \Phi_{1R}},
\;\; {\cal J}_L^{\pm} = 
\frac{1}{2\pi a_0}\; \re^{\pm \ri \sqrt{8\pi} \bar \Phi_{1L}}.
\label{bosophi1}
\eea
The staggered part of a SU(2) spin density can also be expressed in terms of 
the bosonic $\bar \Phi_1$ and its dual field $\bar \Theta_1$
by (see Ref.\cite{shelton} or chapter 13 of Ref. \cite{book1}):
\bea
{\cal N}^z &=& -\frac{\lambda}{\pi a_0} \sin\left(\sqrt{2\pi}\bar\Phi_1\right)
\nonumber\\
{\cal N}^{\pm} &=& \frac{\lambda}{\pi a_0} 
\re^{\pm \ri \sqrt{2\pi} \bar \Theta_1}
\label{stagbosophi1}
\eea
where $\lambda$ is non-universal constant related to 
the band gap for the charge excitations of the underlying  
Hubbard model.
The WZNW primary operator $g$ transforming according to 
the fundamental representation then writes: 
\be 
g = \frac{\pi}{\lambda \sqrt{2}}\left(\epsilon + 
\vec {\cal N} \cdot \vec{\sigma}\right)
\label{prim}
\ee
where $\epsilon = \lambda \cos(\sqrt{2\pi}\bar\Phi_1)/\pi a_0$
is the dimerization operator.
With the bosonic representation (\ref{bosophi1}, \ref{stagbosophi1}),
one can check that the primary field $g$ (\ref{prim}) verifies the following
OPE: 
\bea 
{\cal J}_L^a\left(z\right) g\left(\omega, \bar \omega\right) &\sim& 
-\frac{1}{2\pi\left(z-\omega\right)} \left(\frac{\sigma^a}{2}\right) 
g\left(\omega, \bar \omega\right) \\ 
{\cal J}_R^a\left(\bar z\right) g\left(\omega, \bar \omega\right) &\sim& 
\frac{1}{2\pi\left(\bar z-\bar \omega\right)} g\left(\omega, \bar \omega\right) 
\left(\frac{\sigma^a}{2}\right)
\eea
which corresponds (up to the normalization factor $1/2\pi$) to the 
definition of a WZNW primary operator of Ref. \cite{dms} (chapter 15).

\subsubsection{Representation of the spin currents}

We begin our analysis by considering the SU(2)$_1$ spin 
currents ${\bf J}_{0R,L}$ of the central chain. Using the 
canonical transformation (\ref{can1}, \ref{can2}) with 
the special value (\ref{th2}), we obtain: 
\bea
\varphi_R &=& -\bar \Phi_{1L} + \sqrt{2} \bar \Phi_{2R} \nonumber\\
\varphi_L &=& \sqrt{2} \bar \Phi_{2L} - \bar \Phi_{1R}
\label{transphichir}
\eea
from which together with Eq. (\ref{boso1}) we deduce the 
transformation of the spin currents
of the central chain at the Toulouse point:
\bea
J_{0R,L}^z &=& - {\cal J}_{L,R}^z + \frac{1}{\sqrt \pi} \partial_x
\bar \Phi_{2R,L}\nonumber\\ 
J_{0R}^{\pm} &=& {\cal J}_L^{\pm} 
\re^{\mp \ri \sqrt{16\pi} \bar\Phi_{2R}}\nonumber\\ 
J_{0L}^{\pm} &=& {\cal J}_R^{\pm} 
\re^{\pm \ri \sqrt{16\pi} \bar\Phi_{2L}}
\label{spincentrcur}
\eea
where we have used (\ref{bosophi1}) to express the spin currents
of the central chain in terms of the ``physical'' current ${\cal J}^a$.
Using the correspondence (\ref{psi}), 
$\partial_x \bar \Phi_{2R,L}$ can be related to
the Majorana fermions $\eta_{R,L}$ and $\zeta_{R,L}$:
\be
\partial_x \bar \Phi_{2R,L} = \ri \sqrt \pi \eta_{R,L} \zeta_{R,L}.
\label{parphi2rl}
\ee
Consequently, at the Toulouse point, we deduce the expression of
the z-part of the spin current of the central chain:
\bea
J_{0R}^z &=& - {\cal J}_{L}^z + \ri \eta_R \zeta_R \nonumber\\ 
J_{0L}^z &=& - {\cal J}_{R}^z + \ri \eta_L \zeta_L.
\label{centrcurz1}
\eea
Since the bilinear fields $\eta_{R,L} \zeta_{R,L}$ are 
short-ranged,  we can simplify further (\ref{centrcurz1})
and obtain the following correspondence 
valid for the estimation of the leading contribution 
of correlation functions:
\be
J_{0R,L}^z \sim - {\cal J}_{L,R}^z.
\ee
Notice that the chiralities of the central chain and the physical
spin current are opposite. We shall return to this important 
point when discussing the nature 
of elementary excitations of the model. For the transverse part, 
the bosonic operator $\exp(\pm \ri
\sqrt{16\pi} \bar \Phi_{2R,L})$ has a non-zero expectation value, 
one thus obtains ignoring short-ranged pieces: 
\bea
J_{0R}^{+} &\sim& \gamma_R^{*} {\cal J}_L^{+}\nonumber\\
J_{0R}^{-} &\sim& \gamma_R {\cal J}_L^{-}\nonumber\\
J_{0L}^{+} &\sim& \gamma_L {\cal J}_R^{+}\nonumber\\
J_{0L}^{-} &\sim& \gamma_L^{*} {\cal J}_R^{-}
\label{centrcurpm1}
\eea
where $\gamma_{R,L}$ are defined as:
\be
\langle \re^{\ri
\sqrt{16\pi} \bar \Phi_{2R,L}} \rangle = \gamma_{R,L}.
\label{expec1}
\ee
The actual
value of $\gamma_{R,L}$ is computed in Appendix A where one
can show that $\gamma_{R} = \gamma_{L}=\gamma_m$ and $\gamma_m$
is real. One thus concludes that:
\be
J_{0R,L}^{\pm} \sim \gamma_m {\cal J}_{L,R}^{\pm}. 
\label{centrcurpm2}
\ee

The representation of the SU(2)$_1$ spin currents
of the surface chains (${\bf J}_{aR,L}, a=1,2$)
proceeds in the same way. To each chain $a$, we associate
a massless bosonic field $\varphi_a$ and its dual fields 
$\vartheta_a$ to write
the currents as in Eq. (\ref{boso1}):
\bea
J_{aR}^z &=& \frac{1}{\sqrt{2\pi}}\; \partial_x \varphi_{aR} \;\; ,\;
\; \;\;\; \;
J_{aL}^z = \frac{1}{\sqrt{2\pi}}\; \partial_x \varphi_{aL} \nonumber \\
J_{aR}^{\pm} &=& \frac{1}{2\pi a_0}\; \re^{\mp \ri \sqrt{8\pi} \varphi_{aR}},
\;\; J_{aL}^{\pm} =
\frac{1}{2\pi a_0}\; \re^{\pm \ri \sqrt{8\pi} \varphi_{aL}}.
\label{bosoout}
\eea
To relate these fields to the bosonic field $\Phi$ of
Eq. (\ref{boso3}), let us introduce the symmetric
and antisymmetric combinations:
\be
\varphi_{\pm} = \frac{1}{\sqrt2} \left( \varphi_1 \pm \varphi_2\right),\ \
\vartheta_{\pm} = \frac{1}{\sqrt2} \left( \vartheta_1 \pm \vartheta_2\right),
\label{phipm}
\ee
so that
\bea
J_{aR,L}^z &=& \frac{1}{2\sqrt{\pi}}\left(\partial_x \varphi_{+R,L}
+ \tau_a \partial_x \varphi_{-R,L}\right)\nonumber\\
J_{aR}^{\pm} &=& \frac{1}{2\pi a_0}\; \re^{\mp \ri \sqrt{4\pi} \varphi_{+R}}
\re^{\mp \ri \tau_a\sqrt{4\pi} \varphi_{-R}}\nonumber\\
J_{aL}^{\pm} &=& \frac{1}{2\pi a_0}\; \re^{\pm \ri \sqrt{4\pi} \varphi_{+L}}
\re^{\pm \ri \tau_a\sqrt{4\pi} \varphi_{-L}}
\label{bosoout1}
\eea
where $\tau_1=1$ and $\tau_2=-1$. Moreover,
since we have the relation:
\be
I_{\alpha}^{\pm} = J^{\pm}_{1\alpha}
+ J^{\pm}_{2\alpha}, \;\; \alpha = R,L
\ee
and the expressions (\ref{boso3}) for the bosonization
of the SU(2)$_2$ current ${\bf I}$, we obtain the
following correspondence:
\bea
\varphi_{+R,L} &=& \Phi_{R,L}\nonumber\\
\cos\left(\sqrt{4\pi} \varphi_{-\alpha}\right) &=& \sqrt{\pi a_0} \; \ri
\xi_{\alpha}^3 \kappa , \;\; \alpha=R,L.
\label{cosphim}
\eea
The difference between the SU(2)$_1$ currents of the surface 
chains can also be expressed in terms of the Majorana fermions
$\xi^a, a=0,1,2,3$ (see Eq. (\ref{j1mj2})).
Using Eqs. (\ref{boso2}, \ref{bosoout1}), one thus finds
the identification:
\bea
\partial_x \varphi_{-\alpha} &=& \ri \sqrt \pi \xi_{\alpha}^3 \xi_{\alpha}^0,
\;\; \alpha = R,L \nonumber \\
\sin\left(\sqrt{4\pi} \varphi_{-R}\right) &=& \sqrt{\pi a_0} \; \ri
\xi_{R}^0 \kappa \nonumber \\
\sin\left(\sqrt{4\pi} \varphi_{-L}\right) &=& -\sqrt{\pi a_0} \; \ri
\xi_{L}^0 \kappa .
\label{sinphim}
\eea
With these relations at hand, we can now relate the spin 
currents of the surface chains (\ref{bosoout1}) as a function of the Majorana 
fermions $\xi^0, \xi^3$ and the bosonic field $\Phi$:
\bea
J_{aR,L}^z &=&
\frac{1}{2\sqrt{\pi}}\partial_x \Phi_{R,L}
+ \frac{\ri}{2} \tau_a \xi_{R,L}^3 \xi_{R,L}^0 \nonumber\\
J_{aR}^{\pm} &=& \frac{1}{2\sqrt{\pi a_0}}\; 
\re^{\mp \ri \sqrt{4\pi} \Phi_{R}}
\left(\ri \xi_{R}^3\kappa \pm \tau_a \xi_{R}^0\kappa\right) 
\nonumber\\
J_{aL}^{\pm} &=& \frac{1}{2\sqrt{\pi a_0}}\; 
\re^{\pm \ri \sqrt{4\pi} \Phi_{L}}
\left(\ri \xi_{L}^3\kappa \pm \tau_a \xi_{L}^0\kappa\right). 
\eea
Under the canonical transformation (\ref{can1}, \ref{can2}), 
the Majorana fermions $\xi^0, \xi^3$ remain unchanged while 
the chiral bosonic fields $\Phi_{R,L}$ transform at 
the Toulouse point according to:
\bea
\Phi_{R} &=& \sqrt2 \bar \Phi_{1R}
- \bar \Phi_{2L} \nonumber \\
\Phi_{L} &=& \sqrt2 \bar \Phi_{1L} - \bar \Phi_{2R}.
\label{transphirl}
\eea
Therefore, using the relation (\ref{parphi2rl})
between the Majorana fields $\zeta, \eta$ and the 
bosonic field $\bar \Phi_2$, we obtain for the z-component:
\bea
J_{aR}^z &=& {\cal J}_R^z - \frac{\ri}{2}
\left(\eta_L\zeta_L - \tau_a \xi_R^3 \xi_R^0\right)
\nonumber\\
J_{aL}^z &=& {\cal J}_L^z - \frac{\ri}{2}
\left(\eta_R\zeta_R - \tau_a \xi_L^3 \xi_L^0\right),
\label{outercurz}
\eea
and ignoring the short-ranged contributions, we get:
\be 
J_{aR,L}^z \sim {\cal J}_{R,L}^z .
\label{zpartcurrenouter}
\ee
In the same way, the transverse parts of 
the spin currents of the surface chains simplify as 
follows:
\bea
J_{aR}^{\pm} &=& -\pi a_0 {\cal J}_R^{\pm}\left(
\ri \eta_L\xi_R^3 \pm \zeta_L \xi_R^3 \pm \tau_a \eta_L
\xi_R^0 - \ri \tau_a \zeta_L \xi_R^0\right) \nonumber\\
J_{aL}^{\pm} &=& -\pi a_0 {\cal J}_L^{\pm}\left(
\ri \eta_R\xi_L^3 \pm \zeta_R \xi_L^3 \pm \tau_a \eta_R
\xi_L^0 - \ri \tau_a \zeta_R \xi_L^0\right).
\label{outercurpm}
\eea
Majorana bilinears, in this equation, built from 
$\xi^3_{R,L}$ and $\xi^3_{R,L}$ are short-ranged 
and from the hybridization (\ref{hybrid}), the only 
non-zero vacuum expectation values are:  
\be
{\bar \gamma}_{R,L} = \ri \pi a_0
\langle \xi^3_{R,L} \eta_{L,R} \rangle.
\label{expec2}
\ee
The value of this expectation value is estimated 
in Appendix A and one finds: ${\bar \gamma}_{R} =
{\bar \gamma}_{L} = {\bar \gamma}_{m}$ (${\bar \gamma}_{m}$ 
being real). With the same accuracy than in Eq. (\ref{centrcurpm2}), 
the transverse spin currents of the surface chains are  thus
given by: 
\bea
J_{aR}^{\pm} &\sim& {\cal J}_R^{\pm}\left(
{\bar \gamma}_{m} + \ri a_0 \tau_a\pi\zeta_L \xi_R^0\right) \nonumber\\ 
J_{aL}^{\pm} &\sim& {\cal J}_L^{\pm}\left(
{\bar \gamma}_{m} + \ri a_0 \tau_a\pi
\zeta_R \xi_L^0\right).
\label{outercurpmfin}
\eea

\subsubsection{Representation of the staggered spin fields}

Let us first analyse the staggered spin field of the central 
chain ${\bf n}_0$. As in Eq. (\ref{stagbosophi1}), this field 
can be expressed in terms of the bosonic field $\varphi$
and its dual field $\vartheta$:
\bea
n_0^z &=& -\frac{\lambda}{\pi a_0} \sin\left(\sqrt{2\pi}\varphi\right)
\nonumber\\
n_0^{\pm} &=& \frac{\lambda}{\pi a_0}
\re^{\pm \ri \sqrt{2\pi} \vartheta}.
\label{stagbosovarphi}
\eea
Using the transfomation of the chiral bosonic fields (\ref{transphichir}),
one finds at the Toulouse point:
\bea
\varphi &=& \sqrt 2 \bar \Phi_2 - \bar \Phi_1 \nonumber \\
\vartheta &=& \sqrt 2 \bar \Theta_2 + \bar \Theta_1.
\label{transtouvar}
\eea
We deduce then the following result:
\bea
n_0^z &=& -\frac{\lambda}{\pi a_0} \left(
\sin\left(\sqrt{4\pi}\bar\Phi_2\right) 
\cos\left(\sqrt{2\pi}\bar\Phi_1\right)
-\sin\left(\sqrt{2\pi}\bar\Phi_1\right) 
\cos\left(\sqrt{4\pi}\bar\Phi_2\right)\right)
\nonumber\\
n_0^{\pm} &=& \frac{\lambda}{\pi a_0}
\re^{\pm \ri \sqrt{2\pi} \bar \Theta_1}
\re^{\pm \ri \sqrt{4\pi} \bar \Theta_2}.
\label{stagbosovarphi1}
\eea
One can simplify these expressions
using the correspondence (\ref{psi})
to obtain the relations:
\bea
\cos\left(\sqrt{4\pi}\bar \Phi_2\right) &=&
\ri \pi a_0 \left(\eta_R \eta_L + \zeta_R \zeta_L\right) \nonumber \\
\sin\left(\sqrt{4\pi}\bar \Phi_2\right) &=&
\ri \pi a_0 \left(-\eta_R \zeta_L + \zeta_R \eta_L\right) \nonumber \\
\re^{\pm \ri \sqrt{4\pi} \bar \Theta_2} &=&
\ri \pi a_0 \left(-\eta_R \eta_L \pm \ri \zeta_R \eta_L \pm \ri
\eta_R \zeta_L + \zeta_R \zeta_L\right),
\label{referformphi2}
\eea
from which, we deduce the leading part of the 
representation of the staggered spin fields ${\bf n}_0$ 
associated with the central chain at the Toulouse point: 
\bea
n_0^z &\sim& - \ri \pi a_0 \zeta_R \zeta_L {\cal N}^z \nonumber\\
n_0^{\pm} &\sim& \ri \pi a_0 \zeta_R \zeta_L {\cal N}^{\pm} 
\label{staggeredrepcent}
\eea
where ${\vec {\cal N}}$ corresponds to the staggered magnetization
of the effective S=1/2 Heisenberg spin 
chain (see Eq. (\ref{stagbosophi1})).

The same correspondence for the staggered magnetizations (${\bf n}_a$)
of the surface chains is more difficult to obtain. The reason 
for this is that it 
involves nonlocal strings of the Majorana 
fermions $\xi^a,\; \eta$ and $\zeta$. As we shall see,
the trick will be to relate
all these quantities to the order and disorder parameters
of the underlying Ising models 
present in the problem. The Appendix B gives a review of
the basic definitions for the bosonization of two Ising
models and introduces also the different fields which
will be useful in this subsection.

The staggered part (${\bf n}_a, a=1,2$) of the spin density
of the surface chains can be expressed in terms of 
the bosonic field $\varphi_a$ and its dual $\vartheta_a$ as in 
Eq. (\ref{stagbosovarphi}): 
\bea
n_a^z &=& -\frac{\lambda}{\pi a_0} \sin\left(\sqrt{2\pi}\varphi_a\right)
\nonumber\\
n_a^{\pm} &=& \frac{\lambda}{\pi a_0}
\re^{\pm \ri \sqrt{2\pi} \vartheta_a}.
\label{stagbosophia}
\eea
Introducing the symmetric and antisymmetric combinations (\ref{phipm})
of the bosonic fields, we get:
\bea
n_a^z &=& -\frac{\lambda}{\pi a_0} \left(
\sin\left(\sqrt{\pi}\Phi\right)
\cos\left(\sqrt{\pi}\varphi_-\right)
+\tau_a\cos\left(\sqrt{\pi}\Phi\right)
\sin\left(\sqrt{\pi}\varphi_-\right)\right)
\nonumber\\
n_a^{\pm} &=& \frac{\lambda}{\pi a_0}
\re^{\pm \ri \sqrt{\pi} \Theta}
\re^{\pm \ri \tau_a \sqrt{\pi} \vartheta_-},
\label{stagbosovarphia1}
\eea
where we have used the fact that $\varphi_{+R,L}$ identifies to
the bosonic fields $\Phi_{R,L}$. 
According to the canonical transformation (\ref{transphirl}) at the Toulouse
point, the bosonic fields $\Phi, \Theta$ transform as: 
\bea
\Phi &=& \sqrt 2 \bar \Phi_1 - \bar \Phi_2 \nonumber \\
\Theta &=& \sqrt 2 \bar \Theta_1 + \bar \Theta_2
\label{transtouPhi}
\eea
whereas the fields $\varphi_-, \vartheta_-$ are not 
affected by this transformation.
Using Eqs. (\ref{Isibosigmu03}, \ref{Isibosigmu45}) 
of the Appendix B, one can then relate 
the staggered spin fields ${\bf n}_a$ of 
the surface chains to the diffferent order-disorder
operators of the underlying Ising models of the problem: 
\bea
n_a^z = -\frac{\lambda}{\pi a_0} \left(
\sin\left(\sqrt{2\pi}{\bar \Phi}_1\right)
\left(\mu_5 \mu_4 \mu_3 \mu_0 + \tau_a \sigma_5 \sigma_4
\sigma_3 \sigma_0\right) \right. \nonumber \\
\left. + \cos\left(\sqrt{2\pi}{\bar \Phi}_1\right)
\left(-\sigma_5 \sigma_4 \mu_3 \mu_0 + \tau_a \mu_5 \mu_4
\sigma_3 \sigma_0\right) \right)
\nonumber\\
n_a^{\pm} = \frac{\lambda}{\pi a_0}
\re^{\pm \ri \sqrt{2\pi} {\bar \Theta}_1}
\left(\sigma_5 \mu_4 \sigma_3 \mu_0 - \tau_a 
\mu_5 \sigma_4 \mu_3 \sigma_0 \right. \nonumber \\ 
\left. \pm \ri 
\mu_5 \sigma_4 \sigma_3 \mu_0 \pm \ri \tau_a 
\sigma_5 \mu_4 \mu_3 \sigma_0\right).
\label{stagbosovarphia2}
\eea
The Ising models labelled ``3,5'' are non-critical due to 
the hybridization (\ref{hybrid}) of the Majorana fermions $\eta, \xi^3$ 
at the Toulouse point. 
As shown in Appendix C, the order-disorder operators 
($\sigma_{3,5}, \mu_{3,5}$)
of the corresponding
Ising models 
are short-ranged and have the following vaccum expectation values:
\bea 
\langle \mu_3 \mu_5 \rangle = \langle \sigma_3 \sigma_5 \rangle = \mu \ne 0
\nonumber \\
\langle \sigma_3 \mu_5 \rangle = \langle \mu_3 \sigma_5 \rangle = 0. 
\label{sigmamuvacc}
\eea
Consequently, ignoring short-ranged 
contribution in Eq. (\ref{stagbosovarphia2}), we finally obtain 
the correspondence between 
the staggered magnetization of the surface chains and 
the critical fields at the chiral fixed point: 
\bea
n_a^z &\sim& \mu {\cal N}^z\left(
\mu_4 \mu_0 + \tau_a \sigma_4 \sigma_0\right)
\nonumber\\
n_a^{\pm} &\sim& \mu {\cal N}^{\pm} 
\left(\mu_4 \mu_0 - \tau_a
\sigma_4 \sigma_0 \right). 
\label{nsurfacefin}
\eea
 
Let us summarize our results obtained in this subsection.
Using our Toulouse point approach,
the spin currents and staggered fields of 
the three chains have been expressed in terms 
of the critical fields at the Toulouse point.
Up to short-ranged pieces, the spin currents are given 
by: 
\bea 
J_{0R,L}^z &\sim& - {\cal J}_{L,R}^z \nonumber \\
J_{0R,L}^{\pm} &\sim& \gamma_m {\cal J}_{L,R}^{\pm} \nonumber \\
J_{aR,L}^z &\sim& {\cal J}_{R,L}^z \nonumber \\ 
J_{aR}^{\pm} &\sim& {\cal J}_R^{\pm}\left(
{\bar \gamma}_{m} + \ri a_0 \tau_a\pi\zeta_L \xi_R^0\right) \nonumber\\
J_{aL}^{\pm} &\sim& {\cal J}_L^{\pm}\left(
{\bar \gamma}_{m} + \ri a_0 \tau_a\pi
\zeta_R \xi_L^0\right)
\label{spincurrrepfinal}
\eea
whereas the staggered magnetizations of the different chains
write as follows
\bea 
n_0^z &\sim& - \ri \pi a_0 \zeta_R \zeta_L {\cal N}^z \nonumber\\
n_0^{\pm} &\sim& \ri \pi a_0 \zeta_R \zeta_L {\cal N}^{\pm} \nonumber \\
n_a^z &\sim& \mu {\cal N}^z\left(
\mu_4 \mu_0 + \tau_a \sigma_4 \sigma_0\right)
\nonumber\\
n_a^{\pm} &\sim& \mu {\cal N}^{\pm}
\left(\mu_4 \mu_0 - \tau_a
\sigma_4 \sigma_0 \right).
\label{stagrepfinal}
\eea
These results enable us to 
discuss the spectral properties of the 3-chain cylinder model
as well as the computation 
of the leading 
asymptotics of the spin-spin correlation functions 
at the chiral fixed point as described in 
the following section.

\subsection{Spectral properties}

There are two different kinds of massless elementary excitations
at the Toulouse point:
magnetic excitation described
by the field $\bar{\Phi}_1$, and non-magnetic, singlet, excitations
associated with the two Majorana fermions $\xi^0$ and $\zeta$.
These two groups of excitations are decoupled at the chiral fixed
point. Let us first discuss the nature of the magnetic excitations.

\subsubsection{Magnetic excitations}

The magnetic elementary excitations of the system correspond
to the spinons of the effective S=1/2 Heisenberg chain described by
the bosonic field $\bar{\Phi}_1$.
These spinons appear only in pairs in all physical states and
carry a spin S=1/2.
It is very important to notice that, due to the mixing
of different degrees of freedom
reflected in the canonical transformation (\ref{can1},\ref{can2}),
the ``physical'' spinons, $i.e.$ those defined as
$\sqrt{\pi/2}$-kinks of the field $\bar{\Phi}_1$
should not be misleadingly identified as
the spinons of the central chain.
To get a better understanding of the
structure of the spin excitations at the chiral fixed point,
let us recall the
expressions (\ref{centrcurz1}, \ref{outercurz}) of the currents
$J^z _{1,2}$ and $J^z _0$
in terms of the ``physical'' current
${\cal J}^z = (1/\sqrt{2 \pi}) \p_x\bar{\Phi}_1$ at the Toulouse
point obtained in the previous subsection:
\bea
J^z _{1R(L)} &=& {\cal J}^z _{R(L)} - \frac{\ri}{2}
\left(\eta_{L(R)} \zeta_{L(R)} - \xi^3 _{R(L)} \xi^0 _{R(L)} \right),
\nonumber\\
J^z _{2R(L)} &=& {\cal J}^z _{R(L)} - \frac{\ri}{2}
\left(\eta_{L(R)} \zeta_{L(R)} + \xi^3 _{R(L)} \xi^0 _{R(L)} \right),
\nonumber \\
J^z _{0R(L)} &=& - {\cal J}^z _{L(R)} + \ri \eta_{R(L)} \zeta_{R(L)}.
\label{curr}
\eea
At energies $|\omega| \gg m$, where all Majorana
fields can be considered
as
massless, Eqs. (\ref{curr}) transform back to the
standard definitions of the currents
of the three decoupled chains. In this UV limit, one has a
picture
of three groups of independently propagating spinons. However, in the IR
limit ($|\omega| \ll m$), all Majorana bilinears in
Eq. (\ref{curr}) are characterized by short-ranged correlations (since the
Majorana fermions $\eta$ and $\xi^3$ are massive),
implying that strongly fluctuating
parts of the currents of individual chains are no longer independent.
The spinons of the individual chains turn out to be strongly
correlated in the IR limit to participate at
the formation of a single, physical,
current ${\cal J}^z$.
In fact,
the physical spinon represents a chirally asymmetric,
strongly correlated
state of {\it three} spinons.
Indeed,
consider, for instance, a right-moving
$\sqrt{\pi/2}$-kink of the field $\bar{\Phi}_1$, representing a physical
spinon with the spin projection $S^z = 1/2$. According to the exact
relation
\be
{\cal J}^z _{R(L)} = J^z _{1R(L)} + J^z _{2R(L)} + J^z _{0L(R)}
\label{rel}
\ee
following from Eq. (\ref{curr}), such
an excitation is a combination of two right-moving spinons of the surface
chains, each carrying the spin $S^z = 1/2$, and a left-moving antispinon
of the
central chain, with $S^z = - 1/2$.
The rigidity of such a state is ensured by a finite mass gap in the
$(\eta-\xi^3)$
sector of the model.
Therefore, in physical excitations, two spinons of the surface
chains form a bound state with an antispinon of the central chain.
Otherwisely stated, the polarization of the long-wavelength
magnetic excitations of the surface chains is always the same but
opposite to the polarization of the central chain.
This peculiar structure of the elementary
spin excitations at the chiral fixed point is clearly reflected by the
expression for the spin velocity $u_1$ of the bosonic field
${\bar{\Phi}}_1$: $u_1 = (2v_1 - v_0)/3$.

These excitations correspond to the magnetic excitations of the 
model since a uniform magnetic field $H$ along the z direction
only couples to the ${\bar \Phi}_1$ field.
Indeed,
in the continuum limit, this magnetic field couples to the smooth part
of the total spin density of the three chains:
\be
{\cal H}_H = - \frac{H}{\sqrt{2\pi}} \left(\partial_x \varphi_1 +
\partial_x \varphi_1 + \partial_x \varphi_1\right),
\label{magnecham}
\ee
and performing the Toulouse
transformation (\ref{transtouvar}, \ref{transtouPhi}),
one obtains:
\be
{\cal H}_H = - \frac{H}{\sqrt{2\pi}} \; \partial_x {\bar \Phi}_1.
\label{magnechamtou}
\ee
The uniform susceptibility $\chi$ of the model corresponds thus
to that of a single Heisenberg S=1/2 spin chain
and it is given by (in units of $g\mu_B$):
\be
\chi = \frac{1}{2\pi u_1} = \frac{3}{2\pi \left(2v_1 - v_0\right)}.
\label{unifsuscept}
\ee
Notice that due to the non-trivial structure
of the
spinon at the Toulouse point,
the uniform susceptibility of each chain does not add in the
case of a truly uniform field. In contrast, the susceptibilities
would add coherently if one considers a staggered
magnetic field ($H_a$) in the transverse direction:
$H_a = H(\delta_{a1} + \delta_{a2} -
\delta_{a0})$, $a$ being the chain index.
In that case, one finds the contribution:
\be
\chi_{\perp}^{stag} = \frac{3}{2\pi u_1}.
\label{unifstagsuscept}
\ee

\subsubsection{Pseudo-charge degrees of freedom}

Apart from the non-trivial nature of the spinon, the chiral
fixed point manifests
itself in the existence of massless Z$_2$ singlet excitations.
A first non-magnetic excitation stems from the Majorana fermion
$\xi^0$ that decouples from the rest of the spectrum.
This Majorana fermion describes collective
excitations of singlet pairs
formed on the surface chains.
The nature of the other massless
Majorana fermion $\zeta$ at the chiral fixed point
is less transparent: It is a highly nonlocal
object when expressed
in terms of the original spin operators.
A simple way to understand the role
of these singlet excitations is to combine the
Majorana $\xi^0$ and $\zeta$ fields
into a single Dirac $\Psi_c$
fermion and then bosonize it with the introduction of
a bosonic field ${\tilde \Phi}_c$:
\bea
\Psi_{cR} &=& \frac{\xi^0_R + \ri \zeta_R}{\sqrt{2}}
= \frac{1}{\sqrt{2\pi a_0}}\re^{\ri \sqrt{4\pi} {\tilde \Phi}_{cR}}
\nonumber \\
\Psi_{cL} &=& \frac{\xi^0_L + \ri \zeta_L}{\sqrt{2}}
= \frac{1}{\sqrt{2\pi a_0}}\re^{-\ri \sqrt{4\pi} {\tilde \Phi}_{cL}}.
\label{pseudodef}
\eea
The corresponding
massless
bosonic field ${\tilde \Phi}_c$ ressembles
the scalar field describing the charge
degrees of freedom in the one-dimensional SU(2)
Hubbard model away from half-filling
in the limit  $U=\infty$.
This analogy can be seen by looking at the representation
(\ref{stagrepfinal}) of the staggered fields at the Toulouse point.
Using the bosonization result of two Ising models (\ref{Isibosigmu}),
the total staggered magnetization ${\bf n}_{+}$ of the surface chains
reads as follows in terms of the bosonic field  ${\tilde \Phi}_c$:
\bea
{\bf n}_{+} = {\bf n}_{1} + {\bf n}_{2}
            &\sim& 2 \mu \mu_4 \mu_0 {\vec {\cal N}} \nonumber \\
            &\sim& \cos\left(\sqrt{\pi} {\tilde \Phi}_c \right)
{\vec {\cal N}}.
\label{hubbardrepuinf}
\eea
This bosonized version of ${\bf n}_{+}$ is
reminiscent of the staggered magnetization of the Hubbard
model away from half-filling at $U=\infty$ ($K_c =1/2$ in that
case, see for instance
Refs.\cite{book,book1}).
Apart from a velocity anisotropy ($v_0 \ne u_2$),
the fermionic fields $\xi^0$ and $\zeta$ plays the role of the
charge degrees of freedom in the corresponding Hubbard model.
We shall hence
refer to the gapless Majorana fermions $\xi^0$ and $\zeta$ as
``pseudo-charge'' excitations which account for
the central charge $c=1$ at the IR fixed point.
The spin and pseudo-charge separation is already
manifest at the Toulouse point in the Hamiltonian
(\ref{hfintot}) since
these fermions fields are decoupled from the magnetic excitations
described by the bosonic field ${\bar \Phi}_1$.
This separation of the different modes will also be clearly seen in
the expression of the 
leading asymptotics of correlations functions
that we now estimate.

\subsection{Spin-spin correlation functions at the chiral fixed point}

We are in position to compute the leading 
asymptotics of the spin-spin correlation functions of
the 3-chain cylinder model
at the chiral fixed point. Let us begin 
our analysis by considering the spin-spin correlation functions
between the central spins.

\subsubsection{Correlation functions between the central spins}

Since the SU(2) symmetry is broken to an U(1) symmetry in 
our Toulouse point approach ($g_{\parallel} \ne g_{\perp}$),
we shall discriminate between the z and perpendicular 
components of the correlation function: 
\bea
\langle S_0^z\left(x,\tau\right) S_0^z\left(0,0\right) \rangle =
\langle J_{0R}^z\left(x,\tau\right) J_{0R}^z\left(0,0\right) \rangle +
\left(R\rightarrow L\right) \nonumber \\
+
\left(-1\right)^{x/a_0} 
\langle n_0^z\left(x,\tau\right) n_0^z\left(0,0\right) \rangle,
\label{spincorzmiddle}
\eea
\bea
\langle S_0^{+}\left(x,\tau\right) S_0^{-}\left(0,0\right) \rangle =
\langle J_{0R}^{+}\left(x,\tau\right) J_{0R}^{-}\left(0,0\right) \rangle +
\left(R\rightarrow L\right) \nonumber \\
+
\left(-1\right)^{x/a_0} \langle n_0^{+}\left(x,\tau\right)
n_0^{-}\left(0,0\right) \rangle.
\label{spincorpmmiddle}
\eea
Using the representation (\ref{spincurrrepfinal}, \ref{stagrepfinal}) of 
the uniform and staggered parts of the spin density of 
the central chain at the Toulouse point,
we get the following leading behaviours:
\bea
\langle S_0^z\left(x,\tau\right) S_0^z\left(0,0\right) \rangle \sim 
\langle {\cal J}_{R}^z\left(x,\tau\right) 
{\cal J}_{R}^z\left(0,0\right) \rangle +
\left(R\rightarrow L\right) \nonumber \\
+
\left(-1\right)^{x/a_0} \pi^2 a_0^2
\langle {\cal N}^z\left(x,\tau\right) {\cal N}^z\left(0,0\right) \rangle
\langle \zeta_R\left(x,\tau\right) \zeta_R\left(0,0\right) \rangle
\langle \zeta_L\left(x,\tau\right) \zeta_L\left(0,0\right) \rangle
\label{spincorzmiddlenext}
\eea
and 
\bea
\langle S_0^{+}\left(x,\tau\right) S_0^{-}\left(0,0\right) \rangle \sim 
\gamma_m^2 \langle {\cal J}_{R}^{+}\left(x,\tau\right) 
{\cal J}_{R}^{-}\left(0,0\right) \rangle +
\left(R\rightarrow L\right) \nonumber \\
+
\left(-1\right)^{x/a_0} 
\pi^2 a_0^2
\langle {\cal N}^+\left(x,\tau\right) {\cal N}^-\left(0,0\right) \rangle
\langle \zeta_R\left(x,\tau\right) \zeta_R\left(0,0\right) \rangle
\langle \zeta_L\left(x,\tau\right) \zeta_L\left(0,0\right) \rangle.
\label{spincorpmmiddlenext}
\eea
Accordingly, the estimation of these correlations simply reduces to 
the computation of two-point functions of a massless bosonic field 
${\bar \Phi}_1$ and a massless Majorana fermion $\zeta$. 
The final expressions
for the leading
asymptotics of the correlation functions
between spins in the central chain at the chiral fixed point are thus
\bea
\langle S_0^z\left(x,\tau\right) S_0^z\left(0,0\right) \rangle &\sim&
- \frac{1}{8\pi^2}
\left(\frac{1}{\left(x + \ri u_1 \tau\right)^2} +
\frac{1}{\left(x - \ri u_1 \tau\right)^2} \right) \nonumber\\
&+& \left(-1\right)^{x/a_0} \; \frac{\lambda^2 a_0}{8\pi^2}\;
\frac{1}{\left(x^2+u_2^2\tau^2\right)}
\frac{1}{\left(x^2+u_1^2\tau^2\right)^{1/2}},
\label{spincorzfinmiddle}
\eea
\bea
\langle S_0^{+}\left(x,\tau\right) S_0^{-}\left(0,0\right) \rangle &\sim&
- \frac{\gamma_{m}^2 }{4\pi^2}  \;
\left(\frac{1}{\left(x + \ri u_1 \tau\right)^2} +
\frac{1}{\left(x - \ri u_1 \tau\right)^2} \right) \nonumber \\
&+& \left(-1\right)^{x/a_0} \; \frac{\lambda^2 a_0}{4\pi^2}\;
\frac{1}{\left(x^2+u_2^2\tau^2\right)}
\frac{1}{\left(x^2+u_1^2\tau^2\right)^{1/2}}.
\label{spincorpmfinmiddle}
\eea

\subsubsection{Correlation functions between spins of the surface chains}

Let us now compute the spin-spin correlation functions between 
the surface chains:
\bea
\langle S^z_a\left(x,\tau\right) S^z_b\left(0,0\right) \rangle =
\langle J_{aR}^z\left(x,\tau\right) J_{bR}^z\left(0,0\right) \rangle +
\left(R\rightarrow L\right) +
\left(-1\right)^{x/a_0} 
\langle n^z_a\left(x,\tau\right) n^z_b\left(0,0\right) \rangle,
\label{spincorzout}
\eea
\bea
\langle S^{+}_a\left(x,\tau\right) S^{-}_b\left(0,0\right) \rangle =
\langle J_{aR}^{+}\left(x,\tau\right) J_{bR}^{-}\left(0,0\right) \rangle +
\left(R\rightarrow L\right) +
\left(-1\right)^{x/a_0} \langle n^{+}_a\left(x,\tau\right)
n^{-}_b\left(0,0\right) \rangle.
\label{spincorpmout}
\eea
where $a,b=(1,2)$ is the chain index.
We shall follow the same route as 
for the central chain 
by using the expression of the uniform and staggered 
spin fields (\ref{spincurrrepfinal}, \ref{stagrepfinal}) 
at the Toulouse point: 
\bea
\langle S_a^z\left(x,\tau\right) S_b^z\left(0,0\right) \rangle \sim
\langle {\cal J}_{R}^z\left(x,\tau\right)
{\cal J}_{R}^z\left(0,0\right) \rangle +
\left(R\rightarrow L\right) \nonumber \\
+
\left(-1\right)^{x/a_0} \mu^2
\langle {\cal N}^z\left(x,\tau\right) {\cal N}^z\left(0,0\right) \rangle
\nonumber \\
\left(
\langle \mu_4\left(x,\tau\right) \mu_4\left(0,0\right) \rangle
\langle \mu_0\left(x,\tau\right) \mu_0\left(0,0\right) \rangle
+ \tau_a \tau_b 
\langle \sigma_4\left(x,\tau\right) \sigma_4\left(0,0\right) \rangle
\langle \sigma_0\left(x,\tau\right) \sigma_0\left(0,0\right) \rangle
\right)
\label{spincorzoutnext}
\eea
where we recall that $\tau_1=1,\tau_2 =-1$
and the perpendicular part reads 
\bea
\langle S_a^{+}\left(x,\tau\right) S_b^{-}\left(0,0\right) \rangle \sim 
\langle {\cal J}_{R}^{+}\left(x,\tau\right)
{\cal J}_{R}^{-}\left(0,0\right) \rangle
\left({\bar \gamma}_m^2 + \tau_a \tau_b a_0^2
\langle \zeta_L\left(x,\tau\right) \zeta_L\left(0,0\right) \rangle
\langle \xi^0_R\left(x,\tau\right) \xi^0_R\left(0,0\right) \rangle
\right)
\nonumber \\
+\left(R,L\right)\rightarrow \left(L,R\right)
+
\left(-1\right)^{x/a_0}
\mu^2 
\langle {\cal N}^+\left(x,\tau\right) {\cal N}^-\left(0,0\right) \rangle
\nonumber \\
\left(
\langle \mu_4\left(x,\tau\right) \mu_4\left(0,0\right) \rangle
\langle \mu_0\left(x,\tau\right) \mu_0\left(0,0\right) \rangle
+ \tau_a \tau_b
\langle \sigma_4\left(x,\tau\right) \sigma_4\left(0,0\right) \rangle
\langle \sigma_0\left(x,\tau\right) \sigma_0\left(0,0\right) \rangle
\right).
\label{spincorpmoutnext}
\eea
The Ising models labelled by ``0'' and ``4'' 
associated with respectively the Majorana fermion $\xi^0$ 
and $\zeta$
are decoupled and
critical at the chiral fixed point.
The correlation functions of the order and disorder operators 
$\sigma_{0,4}, \mu_{0,4}$
are known exactly. In the 
long time and long distance limit they are given by (see Appendix B): 
\bea
\langle \mu_0\left(x,\tau\right) \mu_0\left(0,0\right)\rangle &=&
\langle \sigma_0\left(x,\tau\right) 
\sigma_0\left(0,0\right)\rangle  \nonumber \\
 &\sim&\frac{a_0^{1/4}}{(x^2 + v_1^2
\tau^2)^{1/8}},
\label{sigmucor0}
\eea
\bea
\langle \mu_4\left(x,\tau\right) \mu_4\left(0,0\right)\rangle &=&
\langle \sigma_4\left(x,\tau\right) \sigma_4\left(0,0\right)\rangle  
\nonumber \\
 &\sim&\frac{a_0^{1/4}}{(x^2 + u_2^2
\tau^2)^{1/8}}.
\label{sigmucor4}
\eea
This enables us to obtain
the result for the asymptotics of 
the correlation functions between spins
of the surface chains:
\bea
\langle S^z_a\left(x,\tau\right) S^z_b\left(0,0\right) \rangle \sim
 -\frac{1}{8\pi^2}
\left(\frac{1}{\left(x + \ri u_1 \tau\right)^2} +
\frac{1}{\left(x - \ri u_1 \tau\right)^2} \right) \nonumber \\
+ (-1)^{x/a_0}\delta_{ab}
\frac{\mu^2\lambda^2}{\pi^2 \sqrt a_0} \frac{1}{\left(x^2 +
u_1^2\tau^2\right)^{1/2}} \frac{1}{\left(x^2 + v_1^2
\tau^2\right)^{1/8}}\frac{1}{\left(x^2 + u_2^2
\tau^2\right)^{1/8}},
\label{spincorzoutfin}
\eea
\bea
\langle S^{+}_a\left(x,\tau\right) S^{-}_b\left(0,0\right) \rangle \sim
-\frac{{\bar \gamma}_{m}^2}{4\pi^2} 
\left(\frac{1}{\left(x + \ri u_1 \tau\right)^2}
+
\frac{1}{\left(x - \ri u_1 \tau\right)^2}\right)
\nonumber \\
- \frac{a_0^2 \tau_a \tau_b}{16 \pi^4}
\left[
\frac{1}{\left(x - \ri u_1 \tau\right)^2} 
\frac{1}{\left(x - \ri v_1 \tau\right)}
\frac{1}{\left(x + \ri u_2 \tau\right)}
+ 
\frac{1}{\left(x + \ri u_1 \tau\right)^2} 
\frac{1}{\left(x + \ri v_1 \tau\right)} 
\frac{1}{\left(x - \ri u_2 \tau\right)}
\right] \nonumber \\
+ (-1)^{x/a_0} \delta_{ab}
\frac{2 \mu^2 \lambda^2}{\pi^2 \sqrt a_0} \frac{1}{\left(x^2 +
u_1^2\tau^2\right)^{1/2} } \frac{1}{\left(x^2 + v_1^2
\tau^2\right)^{1/8}}\frac{1}{\left(x^2 + u_2^2
\tau^2\right)^{1/8}}.
\label{spincorpmoutfin}
\eea
Here again, the leading contribution of the uniform 
part of the correlation functions is that of a single 
S=1/2 Heisenberg spin chain. The other critical modes 
manifest themselves in subleading contributions in the uniform part
and directly in the staggered part.
In particular, the contributions of the Majorana fermions 
$\xi^0$ and $\zeta$ in the staggered part show up through
their corresponding order and disorder Ising operators.

\subsubsection{Correlation functions between spins of 
the surface chains and of the central chain}

The last spin-spin correlation functions to estimate 
at the chiral fixed point
consists in correlation between a spin of the surface 
chain ${\bf S}_a$ ($a=1,2$) and a spin of the central chain ${\bf S}_0$.
Using the representation (\ref{spincurrrepfinal}, \ref{stagrepfinal})
of the different 
spin fields at the Toulouse point, 
we obtain following the same route as for the previous 
correlation functions:
\bea
\langle S_a^z\left(x,\tau\right) S_0^z\left(0,0\right) \rangle \sim
 \frac{1}{8\pi^2}
\left(\frac{1}{\left(x + \ri u_1 \tau\right)^2} +
\frac{1}{\left(x - \ri u_1 \tau\right)^2} \right), 
\label{spincorzmiddlesurf}
\eea
\bea
\langle S_a^{+}\left(x,\tau\right) S_0^{-}
\left(0,0\right) \rangle \sim
-\frac{{\bar \gamma}_{m}\gamma_m}{4\pi^2}
\left(\frac{1}{\left(x + \ri u_1 \tau\right)^2}
+
\frac{1}{\left(x - \ri u_1 \tau\right)^2}\right).
\label{spincorpmmiddlesurf}
\eea
It is interesting to notice that
the staggered parts of these correlation functions 
are short-ranged at the chiral fixed point.
This result stems from the very special structure of the spectrum 
of the Hamiltonian at the Toulouse point discussed above.

We end this section by estimating the NMR 
relaxation rate $1/ T_1$ at low temperature.
The slowest correlation 
functions of the model at the chiral fixed point correspond to 
the staggered correlations between the spins of the
surface chains (see Eqs. (\ref{spincorzoutfin}, 
\ref{spincorpmoutfin})).
We emphasize that the exponents occuring in these
correlation functions
are {\it universal} and
characterize a new universality class
in spin ladders. We deduce from these correlation 
function (\ref{spincorzoutfin},
\ref{spincorpmoutfin}) the low-temperature dependence of the NMR 
relaxation rate:
\be 
\frac{1}{T_1} \sim \sqrt{T} 
\label{nmr}
\ee 
in contrast with the S=1/2 Heisenberg 
spin chain where $1/ T_1 \sim \;  \rm const$\cite{book,book1}.
Finally, we remark that the result (\ref{nmr})
corresponds to the leading behaviour of the 
NMR relaxation rate of the one-dimensional 
Hubbard model away from half-filling at $U=\infty$\cite{book,book1}.  
The main reason for this stems from the presence of the pseudo-charge
degrees of freedom $\xi^0$, $\zeta$ in the correlation function 
(\ref{spincorzoutfin},
\ref{spincorpmoutfin})
that enter through their corresponding
Ising order-disorder operators ($\sigma_{0,4},
\mu_{0,4}$) with scaling dimension $1/8$.

\section{Stability of the chiral fixed point of the 3-chain cylinder model}

In Sections IV and V, we have carefully analysed the
Hamiltonian consisting in only current-current interaction
describing two marginally coupled SU(2)$_2$ and 
SU(2)$_1$ WZNW models.
The model with 
only current-current interaction has been solved 
exactly using a Toulouse point 
approach. 
At the strong coupling fixed point, 
this Hamiltonian belongs to the chiral stabilized liquids universality 
class with central charge $c=2$. The critical fields at
the chiral fixed point consist of two decoupled gapless
modes: a magnetic sector described by an effective S=1/2
Heisenberg chain associated with the bosonic field 
${\bar \Phi}_1$ and the pseudo-charge degrees of freedom
stemming from the two Majorana fermions $\xi^0$ and $\zeta$.
Moreover, the Toulouse point solution captures 
all universal properties of the chiral fixed point
as shown in Appendix D.

The next step of our approach is to analyse the stability 
of this chiral fixed point.
The non-perturbative basis provided by the Toulouse
point approach enables us to investigate the stability
of the chiral fixed point under several operators 
such as backscattering contributions.
In particular, we shall  first investigate the effect of the interchain
backscattering operator to 
deduce the nature of the 
low energy physics of the 3-chain cylinder model in the vinicity of 
the line ${\bar g} = 0$.

\subsection{Effect of the interchain 
backscattering perturbation at the chiral fixed point}

Let us begin by studying the effect of the interchain coupling 
in the 3-chain cylinder model in the vinicity of the chiral fixed point: 
\be 
{\cal H}_{b} = {\tilde g}\; {\bf n}_0 \cdot {\bf n}_+ . 
\label{intercback}
\ee
From Eq. (\ref{barecoupmod1}), we see that
the coupling constant ${\tilde g}$ is 
independent from the coupling $g$ of the current-current interaction.
One can thus always suppose that  ${\tilde g}$ is small to investigate 
the backscattering term (\ref{intercback}) as a weak perturbation 
at the chiral fixed point: $|{\tilde g}| \ll 1$.
Using the bosonized form of the staggered spin fields 
given by Eqs. (\ref{stagbosovarphi}, \ref{stagbosophia}), 
we have: 
\bea 
{\cal H}_{b} = \frac{2 {\tilde g} \lambda^2}{\pi^2 a_0^2} 
\left(\cos\left(\sqrt{\pi} \vartheta_{-}\right)
\cos\left(\sqrt{2\pi} \vartheta - \sqrt{\pi} \Theta\right)
+ \cos\left(\sqrt{\pi} \varphi_{-}\right)
\sin\left(\sqrt{2\pi} \varphi\right)
\sin\left(\sqrt{\pi} \Phi\right)\right)
\label{interbos}
\eea
so that in terms of the different fields at the Toulouse 
point, the backscattering term (\ref{intercback})  reads
as follows: 
\bea 
{\cal H}_{b} = \frac{2 {\tilde g} \lambda^2}{\pi^2 a_0^2}
\left(\cos\left(\sqrt{\pi} \vartheta_{-}\right)
\cos\left(\sqrt{\pi} {\bar \Theta}_2\right)
-\frac{1}{2} \cos\left(\sqrt{\pi} \varphi_{-}\right)
\cos\left(\sqrt{\pi} {\bar \Phi}_2\right) \right. \nonumber \\
\left. 
+ \frac{1}{2} \cos\left(\sqrt{\pi} \varphi_{-}\right)
\cos\left(3\sqrt{\pi} {\bar \Phi}_2 -\sqrt{8\pi} 
{\bar \Phi}_1\right) \right).
\label{interbostous}
\eea
Using  the Ising dictionnary (see Appendix B),
it can be shown that the contribution in 
the magnetic sector of Eq. (\ref{interbostous}) 
is irrelevant.
The backscattering term affects mostly the
pseudo-charge sector. Indeed, using the results (\ref{proffappendfin})
from the structure of the massive modes at the Toulouse point,
one finds the following estimate at the chiral fixed point: 
\be 
{\cal H}_{b} \sim \frac{\lambda^2 
\mu {\tilde g}}{\pi^2 a_0^2} \mu_0 \mu_4 .
\label{interfin}
\ee
We thus conclude that
the backscattering operator opens a gap, $\Delta_c$, in the
pseudo-charge sector but has no
effect on the magnetic (spinons) excitations.
Standard scaling arguments give an
estimate of the mass gap since 
the operator (\ref{interfin}) has scaling dimension 
$1/4$ and thus $\Delta_c \sim {\tilde g}^{4/7}$.
The chiral fixed point is thus unstable in the far IR limit,
and one expects that the system will flow to
the $c=1$ fixed point
of the standard three-leg spin ladder\cite{arri}.
Of course, the very applicability of the perturbative approach
to the chiral fixed point requires
that $\Delta_c \ll m$,
the condition which can always be satisfied for 
sufficiently small ${\tilde g}$.
Under this condition, there exists an
intermediate but still low-energy region
$\Delta_c \ll E \ll m$ where
the $c=2$ behaviour caused by frustration is dominant.
The physics in this region is
universal and cannot be
understood without having recourse to the chiral
fixed point.
At lower energies, $E \ll \Delta_c$, the system will eventually
cross over to the conventional
critical $c=1$ behaviour.

\subsection{Effect of a weak interaction between 
surface chains at the chiral fixed point}

In this subsection, we investigate the stability of the 
chiral fixed point against a weak interaction between 
the spins of the surface chains: 
\be
{\cal O}_{12} = g_{12} {\bf S}_1 \cdot {\bf S}_2
\simeq g_{12} \left({\bf n}_1 \cdot {\bf n}_2 + 
{\bf J}_1 \cdot {\bf J}_2 \right).
\label{boundaryint}
\ee
Such a term accounts for periodic transverse boundary 
condition in the 3-chain cylinder model and may
change drastically the physics of the chiral fixed point.
In Eq. (\ref{boundaryint}), the coupling constant $g_{12}$
is independent of the coupling $g$ of the current-current
interaction of the model so that on can always consider 
${\cal O}_{12}$ as a weak perturbation at the chiral fixed 
point: $|g_{12}| \ll 1$.

Let us first analyse the ${\bf n}_1 \cdot {\bf n}_2$ contribution 
at the chiral fixed point.
Using the bosonized form (\ref{stagbosophia}) of the staggered magnetizations
of the surface chains and the Toulouse basis (\ref{transtouPhi}), one 
finds:
\be 
{\bf n}_1 \cdot {\bf n}_2 = \frac{\lambda^2}{2 \pi^2 a_0^2}
\left[ \cos\left(\sqrt{4\pi} \varphi_-\right) + 
2 \cos\left(\sqrt{4\pi} \vartheta_-\right)
- \cos\left(\sqrt{8\pi}{\bar \Phi}_1 - \sqrt{4\pi}{\bar \Phi}_2 \right)
\right].
\ee
Using the refermionization 
formula (\ref{cosphim}, \ref{sinphim}, \ref{referformphi2}),
this operator can be expressed in terms of the different fields 
occuring in the Toulouse point solution:
\bea 
{\bf n}_1 \cdot {\bf n}_2 = \frac{\ri \lambda^2}{2 \pi a_0}
\left[- \xi^3_R \xi^3_L + 3 \xi^0_R \xi^0_L -\left(\eta_R \eta_L 
+ \zeta_R \zeta_L\right) \cos\left(\sqrt{8\pi}{\bar \Phi}_1\right) 
\right. \nonumber \\
\left.
- \left(\zeta_R \eta_L  
- \eta_R \zeta_L\right) \sin\left(\sqrt{8\pi}{\bar \Phi}_1\right)
\right].
\eea
Since the Majorana fermions $\xi^3$ and $\eta$ are short-ranged
fields , we can drop their contribution so that:
\be 
{\bf n}_1 \cdot {\bf n}_2 \simeq \frac{\ri \lambda^2}{2 \pi a_0}
\left[3 \xi^0_R \xi^0_L - \zeta_R \zeta_L 
\cos\left(\sqrt{8\pi}{\bar \Phi}_1\right) \right].
\label{n1n2fin}
\ee
The two terms in this equation are of different nature.
One the one hand, the first contribution is a relevant perturbation
(with scaling dimension 1) which identifies with the energy
density operator of the Ising model corresponding to the Majorana
fermion $\xi^0$ and drives the latter model out-of criticality.
On the other hand, the second operator in Eq. (\ref{n1n2fin})
is a naively irrelevant contribution with scaling dimension 3
but, as we shall see below, it is crucial to keep it since it couples 
degrees of freedom of different nature: Magnetic ones described 
by the bosonic field ${\bar \Phi}_1$ and a half of the pseudo-charge
degrees of freedom stemming from the Majorana fermion $\zeta_{R,L}$.

The same analysis can be made for the current-current contribution 
in Eq. (\ref{boundaryint}).
In particular, using the results (\ref{outercurz}, \ref{outercurpm}), 
the leading part of this perturbation at the chiral fixed point 
affects mostly the magnetic degrees of freedom:
\be 
{\bf J}_1 \cdot {\bf J}_2 \simeq 2 {\cal J}^z_R {\cal J}^z_L
-\frac{{\bar \gamma}^2_m}{2\pi a_0} \cos\left(\sqrt{8\pi}{\bar \Phi}_1\right)
\label{j1j2nfin}
\ee
which can be rewritten at the SU(2) point (${\bar \gamma}^2_m =1$,
see Appendix A) in a rotationally invariant form:
\be
{\bf J}_1 \cdot {\bf J}_2 \simeq 2 {\vec {\cal J}}_R \cdot 
{\vec {\cal J}}_L .
\label{j1j2fin}
\ee

With the two results (\ref{n1n2fin}, \ref{j1j2fin}) at hand, we are 
now in position 
to investigate the stability of the chiral fixed point upon
switching on a weak interaction between the surface chains.
The effective interaction (\ref{boundaryint}) at the chiral fixed point
writes:
\be 
{\cal O}_{12} \simeq 2 g_{12} {\vec {\cal J}}_R \cdot
{\vec {\cal J}}_L  + 
\frac{g_{12}\ri \lambda^2}{2 \pi a_0}
\left[3 \xi^0_R \xi^0_L - \zeta_R \zeta_L
\cos\left(\sqrt{8\pi}{\bar \Phi}_1\right) \right].
\label{boundaryinteff}
\ee
Due to the first term, the effect of the operator (${\cal O}_{12}$)
on the chiral fixed point strongly depends on the sign of 
the coupling $g_{12}$. 

For an antiferromagnetic interaction (i.e. $g_{12} >0$),
we expect the succession of two transitions in that case. 
A first one (Ising transition), as soon as the coupling 
$g_{12}$ is switched on, 
the model is still critical but with a smaller central charge $c=3/2$
since, as already emphasized, the Majorana fermion $\xi^0$ acquires 
a mass. As a consequence, the Z$_2$ symmetry with respect to the interchange
of the two surface chains is now broken.  
When $g_{12}$ increases, there is a critical value when 
$2 g_{12}$ exceeds the coupling constant $\gamma <0$ of the marginal irrelevant
current-current interaction of the effective S=1/2 Heisenberg 
chain associated with the bosonic field ${\bar \Phi}_1$. 
In that case, the magnetic sector becomes massive and enters in 
a dimerized phase as in the J$_1$-J$_2$ problem\cite{haldane1}.
Moreover, in this massive phase, the bosonic field ${\bar \Phi}_1$
becomes locked and the operator $\cos\left(\sqrt{8\pi}{\bar \Phi}_1\right)$
acquires a non-zero expectation value: 
$\langle \cos\left(\sqrt{8\pi}{\bar \Phi}_1\right)\rangle \ne 0$.
As a consequence,
the magnetic excitation 
affects indirectly the pseudo-charge
degrees of freedom by giving a mass term for the Majorana fermion
$\zeta$ due to the third term in Eq. (\ref{boundaryinteff}):
$ -g_{12}\ri \lambda^2
\langle \cos\left(\sqrt{8\pi}{\bar \Phi}_1\right)\rangle
\zeta_R \zeta_L
/(2 \pi a_0)$.
Therefore, in the dimerized phase, all degrees of freedom are massive.
In summary, the chiral fixed point is unstable against the presence
of an antiferromagnetic interaction between the surface chains.
There are two distinct Ising transitions in the pseudo-charge sector
whereas the spin part spontaneously dimerized for a sufficiently
strong value of the interaction $g_{12}$.

For a ferromagnetic interaction ($g_{12} <0$), 
the Majorana fermion $\xi^0$ still acquires a mass but now 
the bosonic field ${\bar \Phi}_1$ and the Majorana fermion $\zeta$
remain massless excitations so that
the resulting model has central charge $c=3/2$. 
One should note that a {\sl pseudo-charge fractionalization} occurs
in this $c=3/2$ phase: A half of the degrees of freedom, described
by the Majorana fermion $\xi^0$, in 
the pseudo-charge sector becomes massive by an Ising transition 
whereas the other (i.e. the Majorana fermion $\zeta$) is still critical.
A similar phenomenon for the charge degrees of freedom of a one
dimensional model has been reported in Ref.\cite{fabgoners}. 
The fractionalization manifests itself in the spin-spin correlation
functions in this phase.
The asymptotics of the correlation between spins of the central
chain are still given 
by Eqs. (\ref{spincorzfinmiddle}, \ref{spincorpmfinmiddle}).
The only modifications occur in the calculation of the correlation
functions for the surface chains.
For a ferromagnetic interaction ($g_{12} >0$),
one observes from Eq. (\ref{boundaryinteff})
that the mass $m_0$ of the Majorana fermion $\xi^0$ is positive
($m_0 = -3\lambda^2 g_{12}/(2\pi a_0)$) so that the corresponding
Ising model is in its disordered phase: $\langle \sigma_0 \rangle =0$
and $\langle \mu_0 \rangle \ne 0$.
Using Eqs. (\ref{spincorzoutnext}, \ref{spincorpmoutnext}), one 
deduces the following estimate
for the leading asymptotics of the correlation functions:
\bea
\langle S^z_a\left(x,\tau\right) S^z_b\left(0,0\right) \rangle \sim
 -\frac{1}{8\pi^2}
\left(\frac{1}{\left(x + \ri u_1 \tau\right)^2} +
\frac{1}{\left(x - \ri u_1 \tau\right)^2} \right) \nonumber \\
+ (-1)^{x/a_0} A
\frac{1}{\left(x^2 +
u_1^2\tau^2\right)^{1/2}} 
\frac{1}{\left(x^2 + u_2^2
\tau^2\right)^{1/8}},
\label{newspincorzoutfin}
\eea
\bea
\langle S^{+}_a\left(x,\tau\right) S^{-}_b\left(0,0\right) \rangle \sim
-\frac{{\bar \gamma}_{m}^2}{4\pi^2}
\left(\frac{1}{\left(x + \ri u_1 \tau\right)^2}
+
\frac{1}{\left(x - \ri u_1 \tau\right)^2}\right)
\nonumber \\
+ 2 (-1)^{x/a_0} A
\frac{1}{\left(x^2 +
u_1^2\tau^2\right)^{1/2} }
\frac{1}{\left(x^2 + u_2^2
\tau^2\right)^{1/8}}
\label{newspincorpmoutfin}
\eea
where $A$ is a non-universal constant. 
In this $c=3/2$ phase, the slowest correlations still 
correspond to the correlator of the staggered magnetization 
of the surface chains and the low-temperature dependence of 
the NMR relaxation rate is now given by: 
\be 
\frac{1}{T_1} \sim T^{1/4}
\ee
in contrast with the estimate (\ref{nmr}) obtained at the chiral 
fixed point.

\section{Concluding remarks}

In this work, we have studied some specific models which 
display in the long distance limit some CSL physics and 
belongs to a new non-Fermi-Liquid class of fixed points
describing the IR behaviour of one-dimensional interacting
chiral fermions.
The possible realizations of this CSL state, proposed in 
this paper, belong to several topics of the one-dimensional
Quantum Magnetism: the (N+1)-chain cylinder as
a special frustrated spin ladder, its asymmetric doped version 
with a doped central chain as an example of a Luttinger 
liquid in an active environment, and finally the Kondo-Heisenberg 
model with N channels away from half-filling as a generalized 
multichannel Kondo lattice.

The common feature of all these models stems from  
the fact that their Hamiltonians in the continuum limit
have the same structure consisting of two marginally 
coupled SU(2)$_N$ and SU(2)$_1$ WZNW models with only 
current-current interactions. In particular, 
this Hamiltonian decomposes into two commuting {\sl chirally 
asymmetric} parts which transform into each other under the 
time-reversal symmetry. Each part flows to an intermediate
fixed point  where the system as a whole displays critical 
properties characterized by a non-trivial symmetry group 
belonging to the universality class of the CSL state\cite{andrei}.
All the possible lattice realizations are expected to exhibit 
critical properties in the IR limit with, in general, non-integer
central charge.

For the $N=2$ case, by allowing anisotropic interactions in 
spin space, the model of two marginally 
coupled WZNW models can be solved using a Toulouse point approach. It 
provides a direct reading of the spectrum by identifying
the nature of the massive and massless degrees of freedom in 
the IR limit. Moreover, the Toulouse limit solution captures 
all universal properties of the model including the SU(2)
symmetric case. At the strong coupling fixed point, 
the massless degrees of freedom with 
central charge $c=2$ consist of an effective S=1/2
Heisenberg spin chain and two critical Ising models which 
act as pseudo-charge degrees of freedom.
The special nature of the chiral fixed point identified by 
the Toulouse point approach
reveals itself in two important facts for the 3-chain cylinder model.
On the one hand, the magnetic excitations consist of 
a strongly correlated bound-state made by two spinons
of the surface chains and an antispinon of the central chain.
On the other hand, the slowest spin-spin correlation functions 
corresponds to the staggered magnetizations
contribution of the surface chains with an unusual exponent which
characterizes a new universality class in spin ladders
and manifest itself in physical quantity such as the
low-temperature behaviour of the NMR relaxation rate:
$1/T_1 \sim \sqrt T$.

The Toulouse limit solution provides also a non-perturbative
basis to investigate the stability of the chiral fixed point
under the presence of several operators such as backscattering
contributions.
In particular, the interchain backscattering term of the 3-chain
cylinder model destabilizes the chiral fixed point and the 
system will cross over to a fixed point with central charge $c=1$
which is expected to be the one for the non-frustrated 3-leg spin
ladder with open transverse boundary conditions.
However, there still exists an intermediate low-energy region governed 
by the chiral fixed point.
This fixed point becomes also unstable upon 
switching on a weak interaction between the surface chains.
For a ferromagnetic interaction, the system exhibits an interesting
critical phase with a smaller central charge $c=3/2$ where a 
half of the pseudo charge degrees of freedom becomes massive by
an Ising transition and the low temperature 
behaviour of the NMR relaxation rate is now: $1/T_1 \sim T^{1/4}$.
In the case of antiferromagnetic interaction (frustrated periodic
transverse boundary conditions), we expect the succession of two 
transitions with an intermediate $c=3/2$ phase followed by a massive
phase where the magnetic sector is spontaneously dimerized.

Regarding perspectives, it is clearly of interest to further 
explore the phase diagram of the (N+1)-chain cylinder model in
the vinicity of the chiral fixed point.
The experience gained by the Toulouse approach for $N=2$
leads us to expect, though it requires a proof, that the 
intermediate fixed point will become unstable and 
the system may cross over to a fixed point presumably
characterized by a central charge $c=1$. 
A Toulouse limit solution for the general $N$ is thus clearly desirable
to shed light on this problem.
Using a parafermionic description of the SU(2)$_N$ spin current\cite{zamolo2},
one can readily find a Toulouse limit for every $N$
with the decoupling of a bosonic field corresponding to the 
bosonic field ${\bar \Phi}_1$ in the $N=2$ case.
Unfortunately, the other critical and massive degrees of freedom
strongly interact with each other: The resulting effective theory
cannot mapped onto Majorana fermions as in $N=2$ case. 
Otherwisely stated,
when N exceeds two, the Toulouse limit does not correspond to 
a theory of free particles. The simplest case is the $N=4$ case 
which is currently under study.
With this respect, the situation is in close parallel to 
the multichannel Kondo problem since when the number of channels
exceeds two, the Toulouse limit is described by a non-trivial 
field theory\cite{fabrizio,tsvelikto}. 

This analogy leads us to expect that the simple model of a
Luttinger liquid in an antiferromagnetic environment considered in 
this work and the multichannel Kondo-Heisenberg model away from
half-filling should exhibit a non-Fermi-liquid low-temperature 
behaviour with enhanced composite pairing correlations (odd-frequency
pairing) as in the two channel Kondo problem\cite{emery} or
the one-dimensional Kondo lattice\cite{zachar,georges}.
This question in the 
$N=2$ case will be addressed by the Toulouse point approach
in a forthcoming publication\cite{azarial}.
We hope that the chirally stabilized liquid state with all
its physical properties will be observed in numerical simulations
and in further experiments on spin ladder systems.

{\bf Acknowlegdments}

The authors are indepted to A. A. Nersesyan for 
important contributions related to this work.
We would like also to thank
D. Allen and A. O. Gogolin
for very interesting discussions.
One of us (P. L.) would like to thank the 
I.C.T.P. (Trieste) for hospitality during the completion 
of this work.
This paper is dedicated to Meli.

{\bf Note added}   

When this work was completed, we became aware of a work 
by N. Andrei and E. Orignac\cite{orignac} who have also
determined the low energy theory of the one dimensional 
multichannel Kondo-Heisenberg lattice away from half filling
and found that composite order parameters induce the dominant
instabilities for a number of channels less than 4.

\appendix

\section*{A Estimation of $\gamma_m$ and $\bar \gamma_m$}

In this Appendix, we shall compute the constants 
$\gamma_m, {\bar \gamma}_m$ that appear in the 
representation (\ref{spincurrrepfinal}) of the different spin
current fields in 
terms of the current ${\vec {\cal J}}$ of the effective S=1/2 Heisenberg spin
chain at the Toulouse point. 

\subsection{Estimation of $\gamma_m$}

Let us first consider the following vacuum expectation 
values: 
\bea 
\gamma_{R} &=& \langle \exp\left(\pm \ri \sqrt{16\pi} {\bar \Phi}_{2R}
\right) \rangle \nonumber \\
\gamma_{L} &=& \langle \exp\left(\pm \ri \sqrt{16\pi} {\bar \Phi}_{2L}
\right) \rangle .
\label{correspdefapp}
\eea
In the following, we shall show that
$\gamma_R = \gamma_L = 
\gamma_m$, $\gamma_m$ being real and 
expresses in terms of the mass $m$ of the massive modes at
the Toulouse point.

Using the definition (\ref{psi}) of the chiral 
bosonic field ${\bar \Phi}_{2R,L}$, we have:
\bea 
\eta_{R,L} &=& \frac{\kappa}{\sqrt \pi} 
:\cos\left(\sqrt{16\pi} {\bar \Phi}_{2R,L}\right): \nonumber \\
\zeta_{R,L} &=& \pm \frac{\kappa}{\sqrt \pi} 
:\sin\left(\sqrt{16\pi} {\bar \Phi}_{2R,L}\right):
\label{majobosapp}
\eea
where we have used for commodity the normal ordered form 
of an exponential of chiral bosonic fields:
\be 
:\exp\left(\ri \alpha {\bar \Phi}_{2R,L}\right): =
a_0^{-\alpha^2/8\pi} \exp\left(\ri \alpha {\bar \Phi}_{2R,L}\right).
\ee 
The next step is to express the previous exponential bosonic 
fields with scaling dimensions $2$ in terms of the different 
Majorana fields. This can be done by considering the following
OPE from (\ref{majobosapp}): 
\bea 
\eta_R\left(\bar z\right) \eta_R\left(\bar \omega\right) &\sim& 
\frac{1}{2\pi\left(\bar z -\bar \omega\right)} - 
4 \left(\bar z -\bar \omega\right) 
:\left({\bar \partial} {\bar \Phi}_{2R}\right)^2: \nonumber \\
+ \frac{\bar z -\bar \omega}{2\pi} 
:\cos\left(\sqrt{16\pi} {\bar \Phi}_{2R}\right): \nonumber \\ 
\zeta_R\left(\bar z\right) \zeta_R\left(\bar \omega\right) &\sim& 
\frac{1}{2\pi\left(\bar z -\bar \omega\right)} - 
4 \left(\bar z -\bar \omega\right) 
:\left({\bar \partial} {\bar \Phi}_{2R}\right)^2: \nonumber \\
- \frac{\bar z -\bar \omega}{2\pi} 
:\cos\left(\sqrt{16\pi} {\bar \Phi}_{2R}\right):
\label{opemajcosbo}
\eea
where the fields in the second part of the previous equations 
are only function of $\bar \omega$.
On the other hand, since $\eta, \zeta$ are Majorana fermions, 
they satisfy the following OPEs:
\bea 
\eta_R\left(\bar z\right) \eta_R\left(\bar \omega\right) &\sim&
\frac{1}{2\pi\left(\bar z -\bar \omega\right)} -
\left(\bar z -\bar \omega\right)
:\eta_R{\bar \partial} \eta_R: \nonumber \\
\zeta_R\left(\bar z\right) \zeta_R\left(\bar \omega\right) &\sim&
\frac{1}{2\pi\left(\bar z -\bar \omega\right)} -
\left(\bar z -\bar \omega\right)
:\zeta_R{\bar \partial} \zeta_R:. 
\label{opemajgenapp}
\eea
Comparing the two short distance 
expansions (\ref{opemajcosbo}, \ref{opemajgenapp}), we end 
with the equivalence: 
\be 
\cos\left(\sqrt{16\pi} {\bar \Phi}_{2R}\right) = a_0^2 \ri \pi
\left(-\eta_R\partial_x \eta_R + 
\zeta_R\partial_x \zeta_R \right),
\label{gammamajrepright}
\ee
since for instance, in our convention, 
${\bar \partial} \zeta_R = i \partial_x
\zeta_R$ ($z= u_2\tau +\ri x$).
One can do the same computation for the left sector to find:
\be 
\cos\left(\sqrt{16\pi} {\bar \Phi}_{2L}\right) = a_0^2 \ri \pi
\left(\eta_L\partial_x \eta_L -     
\zeta_L\partial_x \zeta_L \right).
\label{gammamajrepleftt}
\ee
In the same way, considering the OPE $\eta_R(\bar z) 
\zeta_R(\bar \omega)$, we get: 
\be 
\sin\left(\sqrt{16\pi} {\bar \Phi}_{2R,L}\right) = \pm
2 \ri \pi a_0^2 \; \partial_x \eta_{R,L} \zeta_{R,L}.
\ee
This proves that $\gamma_{R,L}$ is real since at 
the chiral fixed point $\zeta$ is a free and massless Majorana fermion 
so that $\langle \sin\left(\sqrt{16\pi} {\bar \Phi}_{2R,L}\right) \rangle
= 0$.
Therefore, using Eqs. (\ref{gammamajrepright}, 
\ref{gammamajrepleftt}) and the definition 
(\ref{correspdefapp}),  we obtain the following correspondence: 
\be 
\gamma_{R,L} = \mp a_0^2 \ri \pi 
\left(\langle \eta_{R,L}\partial_x \eta_{R,L} \rangle -
\langle \zeta_{R,L}\partial_x \zeta_{R,L}\rangle \right). 
\label{fincorrespapp}
\ee

The contribution of the massless 
Majorana fermion $\zeta$
is easy to find: 
\bea 
\langle \zeta_{R,L}\partial_x \zeta_{R,L}\rangle &=& 
\int \frac{dk}{2\pi} \int \frac{d\omega}{2\pi} 
\; \frac{\ri k \re^{\ri \omega 0^+}}{\ri \omega \mp u_2 k} \nonumber \\
&=& \mp \frac{\ri}{2\pi} \int_{0}^{\Lambda} dk \; k = 
\mp \frac{\ri \Lambda^2}{4\pi} 
\label{contzeta}
\eea
where we have used a simple cut-off regularization to cure 
the UV singularity.
To estimate $\langle \eta_{R,L}\partial_x \eta_{R,L} \rangle$,
we need to know the Green's function of the Majorana fermion $\eta$
that comes from the massive part of the Hamiltonian at the Toulouse
point describing the hybridization between 
the Majorana fermions $\xi^3$ and $\eta$:
\bea
{\cal H}_{+} &=& \frac{\ri v_1}{2}\xi_L^{3}\partial_x \xi_L^{3}
 -\frac{\ri u_2}{2}
\eta_R\partial_x \eta_R
-\frac{\ri m}{2} \left(\eta_R\xi_L^{3} - \xi_L^{3} \eta_R\right) \nonumber \\
{\cal H}_{-} &=& -\frac{\ri v_1}{2}\xi_R^{3}\partial_x \xi_R^{3}
+\frac{\ri u_2}{2}
\eta_L\partial_x \eta_L
+\frac{\ri m}{2} \left(\xi_R^{3}\eta_L - \eta_L\xi_R^{3}\right).
\label{hybridapp}
\eea
The Green's functions corresponding to ${\cal H}_{+}$
are given by: 
\bea
G_{+}\left(k,\omega\right) = \left(
\begin{array}{lccr}
\langle \eta_R\left(-k,-\omega\right) \eta_R\left(k,\omega\right)
\rangle & \langle \eta_R\left(-k,-\omega\right) \xi_L^3\left(k,\omega\right) 
\rangle \\
\langle \xi^3_L\left(-k,-\omega\right) 
\eta_R\left(k,\omega\right) \rangle & \langle
\xi^3_L\left(-k,-\omega\right) \xi^3_L\left(k,\omega\right) \rangle
\end{array}
\right)
\label{grep}
\eea
with 
\bea
G_{+}^{-1}\left(k,\omega\right) = \left(
\begin{array}{lccr}
\ri \omega - u_2 k &
\ri m \\
-\ri m & 
\ri \omega + v_1 k 
\end{array}
\right),
\label{grepinv}
\eea
whereas the ones associated with ${\cal H}_{-}$ read as 
follows:
\bea
G_{-}\left(k,\omega\right) = \left(
\begin{array}{lccr}
\langle \xi^3_R\left(-k,-\omega\right) \xi^3_R\left(k,\omega\right)
\rangle & \langle \xi^3_R\left(-k,-\omega\right) \eta_L\left(k,\omega\right)
\rangle \\
\langle \eta_L\left(-k,-\omega\right) 
\xi_R^3\left(k,\omega\right) \rangle & \langle
\eta_L\left(-k,-\omega\right) \eta_L\left(k,\omega\right) \rangle
\end{array}
\right)
\label{grem}
\eea
with 
\bea
G_{-}^{-1}\left(k,\omega\right) = \left(
\begin{array}{lccr}
\ri \omega - v_1 k &
-\ri m \\
+\ri m & 
\ri \omega + u_2 k 
\end{array}
\right).
\label{greminv}
\eea
The Majorana Green's function for the $\eta$ field
is thus given by: 
\be 
 \langle
\eta_{R,L}\left(-k,-\omega\right) 
\eta_{R,L}\left(k,\omega\right) \rangle
= - \frac{\ri \omega \pm v_1 k}{\left(\omega \mp \ri k u_1\right)^2 
+ E^2\left(k\right)} 
\ee 
where $E^2(k) = u_{*}^2 k^2 + m^2$ and 
$u_{*} = (v_0 + v_1)/3$.
Integrating over the frequency, we obtain the
following estimate: 
\bea 
\langle \eta_{R,L}\partial_x \eta_{R,L} \rangle &=& 
\mp \frac{\ri u_{*}}{2\pi} \int_{0}^{\Lambda} dk \; 
\frac{k^2}{E\left(k\right)} \nonumber \\
&\simeq& \mp \frac{\ri \Lambda^2}{4\pi} \pm 
\frac{\ri m^2}{4\pi u_{*}^2} \ln\left(\frac{\Lambda u_{*}}{m}\right). 
\label{etaparetaes}
\eea
subsituting (\ref{etaparetaes}) and (\ref{contzeta}) into
(\ref{fincorrespapp}), we find the value of $\gamma_m$ at 
the Toulouse point:
\be 
\gamma_m = \gamma_{R,L} \simeq 
\frac{m^2 a_0^2}{4 u_{*}^2} \ln\left(\frac{\Lambda u_{*}}{m}\right).
\label{gammamest}
\ee

\subsection{Estimation of ${\bar \gamma}_m$}

The next vacuum expectation values to compute is
\be 
{\bar \gamma}_{R,L} = \ri \pi a_0 \langle \xi^3_{R,L} 
\eta_{L,R} \rangle.
\label{gammabarapp}
\ee
Using the previous expression of the 
fermionic Green's function $G_{\pm}$, we have: 
\be
\langle \xi^3_{R,L}\left(-k,-\omega\right) 
 \eta_{L,R}\left(k,\omega\right) \rangle = - 
\frac{\ri m }{\left(\omega \pm \ri k u_1\right)^2
+ E^2\left(k\right)}.
\ee
Integrating over $\omega$, the value of ${\bar \gamma}_{R,L}$
can easily be determined:
\be
{\bar \gamma}_{R,L} = \frac{m a_0}{2} 
\int_{0}^{\Lambda}\;
\frac{dk}{E\left(k\right)}, 
\ee
and thus
\be 
{\bar \gamma}_m = {\bar \gamma}_{R,L} \simeq
\frac{m a_0}{2u_{*}} \ln\left(\frac{\Lambda u_{*}}{m}\right).
\label{gammambarest}
\ee

Let us conclude this appendix by discussing the SU(2) invariance 
of the spin-spin correlation functions between the central spins
and also between the surface chains at the chiral fixed point.
Using the Toulouse point value (\ref{toulouse}): $g_{\parallel}^{*}
= 4\pi u_{*}$
and expressing the values of $\gamma_m, {\bar \gamma}_m$ in 
terms of the ratio ${\bar \kappa} = g_{\perp}/g_{\parallel}$, 
we have: 
\bea 
\gamma_m &\simeq& {\bar \kappa}^2 
\ln\left(\frac{\Lambda a_0}{2 {\bar \kappa}}\right)
\nonumber \\
{\bar \gamma}_m &\simeq& {\bar \kappa}
\ln\left(\frac{\Lambda a_0}{2 {\bar \kappa}}\right).
\label{scalinsu2}
\eea
In the continuum limit, the value of the physical mass 
is fixed. This gives the dependence of $g_{\perp}$ in terms 
of the short distance cut-off $a_0$. The ratio of 
the couplings $g_{\perp}/g_{\parallel}$ is also fixed to 
some physical value ${\bar \kappa}$ which is a RG invariant flow
of the equation (\ref{rgxxz}).
The two physical parameters emerging from the renormalized theory are  
precisely $m$ and ${\bar \kappa}$ apart from the different velocities
of the problem.
At the SU(2) point when ${\bar \kappa} =1$, we must have 
$\gamma_m^2={\bar \gamma}_m^2 =1$ so that the correlation functions
of the central chain (\ref{spincorzfinmiddle}, 
\ref{spincorpmfinmiddle}) 
are rotational invariant. 
Using Eq. (\ref{scalinsu2}), this fixes the value of the 
product $\Lambda a_0$ to $2$.

\section*{B Basic facts about the bosozination of two Ising models}

In this Appendix, we briefly review some basic materials 
on the relation between two Ising models and a free bosonic
field in order to 
fix notations for the computation of the spin-spin correlation 
function between spins of the surface chains (Section V).

It is well known that  
a theory of free massless Majorana fermion 
with central charge $c=1/2$ describes 
the long distance properties of the critical two-dimensional Ising model. 
Two Ising models can be mapped onto the Gaussian model
with central charge $c=1$ (see for a review chapter 12 of 
Ref. \cite{book1}).  
This procedure of doubling the Ising models has been very fruitful 
for the computation of the correlation function of the Ising 
model\cite{zuber,truong,francesco,boyanovsky}, for
the problem of random impurities in the Ising model\cite{shankar},
and also for the calculation of the structure factor of the 
two-leg spin ladder\cite{shelton}. 
This mapping stems from the fact that the two Majorana fields say 
$\psi_{1,2}$ can be combined into a single Dirac fermion $\chi$ 
which can be in turn be expressed in terms of a bosonic field 
$\Phi$ by Abelian bosonization:
\bea 
\chi_R &=& \frac{\psi_{1R} + \ri \psi_{2R}}{\sqrt 2} = 
\frac{1}{\sqrt{2 \pi a_0}} \; \re^{i\sqrt{4\pi} \Phi_R} \nonumber \\ 
\chi_L &=& \frac{\psi_{1L} + \ri \psi_{2L}}{\sqrt 2} = 
\frac{1}{\sqrt{2 \pi a_0}} \; \re^{-i\sqrt{4\pi} \Phi_L}
\label{chiboso}
\eea 
with $[\Phi_R,\Phi_L] = i/4$ to insure the anticommutation relation between 
$\chi_R$ and $\chi_L$.
The Majorana fermions $\psi_{1,2}$ can therefore be expressed 
directly in terms the $\Phi$ field as follows:
\bea
\psi_{1R} &=& \frac{1}{\sqrt{\pi a_0}} \cos\left(\sqrt{4\pi}
\Phi_R\right) \nonumber \\
\psi_{1L} &=& \frac{1}{\sqrt{\pi a_0}} \cos\left(\sqrt{4\pi}
\Phi_L\right) \nonumber \\
\psi_{2R} &=& \frac{1}{\sqrt{\pi a_0}} \sin\left(\sqrt{4\pi}
\Phi_R\right) \nonumber \\
\psi_{2L} &=& -\frac{1}{\sqrt{\pi a_0}} \sin\left(\sqrt{4\pi}
\Phi_L\right).
\label{Isibos1}
\eea
The energy density operator $\epsilon_a = 
i \psi_{aR} \psi_{aL}, a=1,2$, which is a mass term for the 
Majorana fermion and
drives the model out of criticality, can also be written
down as a function of the bosonic field $\Phi$ and its dual $\Theta$:
\bea
\cos\left(\sqrt{4\pi}\Phi\right) &=& \ri \pi a_0
\left(\psi_{1R}\psi_{1L} + \psi_{2R}\psi_{2L}\right) \nonumber \\
\cos\left(\sqrt{4\pi}\Theta\right) &=& \ri \pi a_0
\left(-\psi_{1R}\psi_{1L} + \psi_{2R}\psi_{2L}\right), 
\label{Isibos2}
\eea
and one can also relate all bilinears involving $\psi_{1R,L}$ 
and $\psi_{2R,L}$ to the bosonic field $\Phi$ and its dual $\Theta$:
\bea
\sin\left(\sqrt{4\pi}\Phi\right) &=& \ri \pi a_0
\left(\psi_{2R}\psi_{1L} - \psi_{1R}\psi_{2L}\right) \nonumber \\
\sin\left(\sqrt{4\pi}\Theta\right) &=& \ri \pi a_0
\left(\psi_{2R}\psi_{1L} + \psi_{1R}\psi_{2L}\right),
\label{Isibos3}
\eea
and 
\bea
\partial_x\Phi &=& \ri \sqrt\pi
\left(\psi_{1R}\psi_{2R} + \psi_{1L}\psi_{2L}\right) \nonumber \\
\partial_x\Theta &=& \ri \sqrt\pi
\left(-\psi_{1R}\psi_{2R} + \psi_{1L}\psi_{2L}\right).
\label{Isibos4}
\eea
The Ising model labelled 1, for instance, 
contains apart from the Majorana fields $\psi_{1R,L}$ 
that have conformal dimensions $(1/2,0)$
and
$(0,1/2)$ and the energy density operator, 
two other fields, with conformal dimensions $(1/8,1/8)$:
the order and disorder
operators $\sigma_1$ and $\mu_1$. These fields are non local 
when expressed in terms of the Majorana 
fermion $\psi_{1R,L}$\cite{zuber,truong,boyanovsky}. 
The Kramers-Wannier duality transformation maps the order 
operator $\sigma_1$ onto the disorder field $\mu_1$. In the 
ordered phase ($T < T_c$), $\langle \sigma_1 \rangle 
\ne 0, \langle \mu_1 \rangle =0$ whereas 
in the disorder phase ($T > T_c$), we have 
$\langle \sigma_1 \rangle 
= 0, \langle \mu_1 \rangle  \ne 0$. At $T=T_c$, both 
fields have a zero vacuum expectation value.
At $T=T_c$, the products $\sigma_1 \sigma_2$, $\sigma_1 \mu_2$,
$\mu_1 \mu_2$ and $\mu_1 \sigma_2$ have scaling 
dimension $1/4$ and can be related to the bosonic 
exponentials $\exp(\pm \ri \sqrt \pi \Phi)$, 
$\exp(\pm \ri \sqrt \pi \Theta)$\cite{boyanovsky,shelton}: 
\bea
\mu_1\mu_2 &\sim & \cos \left(\sqrt \pi \Phi\right) \nonumber \\
\sigma_1\sigma_2 &\sim & \sin \left(\sqrt \pi \Phi\right) \nonumber \\
\sigma_1\mu_2 &\sim & \cos \left(\sqrt \pi \Theta\right) \nonumber \\
\mu_1\sigma_2 &\sim & \sin \left(\sqrt \pi \Theta\right),
\label{Isibosigmu}
\eea
this correspondence should also hold at small deviations 
from criticality. These operators are very useful 
for pratical computation since the correlation functions
of the order and disorder fields are known exactly 
even out of criticality. For $T=T_c$, we have the following 
power law in the long distance-long time limit: 
\bea
\langle \mu\left(x,\tau\right) \mu\left(0,0\right)\rangle &=&
\langle \sigma\left(x,\tau\right) \sigma\left(0,0\right)\rangle
\nonumber \\
 &\sim&\frac{1}{(x^2 + v^2
\tau^2)^{1/8}},
\label{Isigmucor}
\eea
$v$ being the 
spin velocity of the Majorana fermion.  
The description of spin chains in terms of order-disorder
parameters of the underlying Ising models has been used 
by Shelton et al.\cite{shelton,nersesyan} for the two-leg spin ladder and
by Allen and S\'en\'echal \cite{allen} for the two-leg zigzag ladder
in the absence of the twist term\cite{nersesyan1}. This 
approach allows a simple description of the excitations 
of the system and is very useful for the computation 
of the correlation functions and the dynamic structure factors
in a massive phase. 

In our problem, we shall use this approach for the computation
of the spin-spin correlation functions of spins of the 
surface chains and for 
the study of the stability of 
the chiral fixed point perturbed by 
some operators. One has to express, at the Toulouse point, 
all the order-disorder operators of the underlying Ising
models. First of all, the bosonic 
field $\varphi_{-} = (\varphi_1 -\varphi_2)/\sqrt 2$ 
is not affected by the canonical transformation (\ref{can1}, \ref{can2}) and 
is built from the two Majorana fermions $\xi^0, \xi^3$.
The two corresponding Ising 
models are labelled ``0'' and ``3'' 
respectively and using Eq. (\ref{Isibosigmu})
with the identification (\ref{cosphim}, \ref{sinphim}), we have: 
\bea
\mu_3\mu_0 &\sim & \cos \left(\sqrt \pi \varphi_{-}\right) \nonumber \\
\sigma_3\sigma_0 &\sim & \sin \left(\sqrt \pi \varphi_{-}\right) \nonumber \\
\sigma_3\mu_0 &\sim& \cos \left(\sqrt \pi \vartheta_{-}\right) \nonumber \\
\mu_3\sigma_0 &\sim& \sin \left(\sqrt \pi \vartheta_{-}\right).
\label{Isibosigmu03}
\eea
A second couple of Ising models (noted ``4'' and ``5'') stems from the two 
Majorana fermions $\zeta, \eta$ respectively and are associated with 
the $\bar \Phi_2$ bosonic field (see Eq. (\ref{psi})). 
In that case, the correspondence (\ref{Isibosigmu}) gives
\bea
\mu_5\mu_4 &\sim& \cos \left(\sqrt \pi \bar \Phi_2 \right) \nonumber \\
\sigma_5\sigma_4 &\sim& \sin \left(\sqrt \pi \bar \Phi_2 \right) \nonumber \\
\sigma_5\mu_4 &\sim& \cos \left(\sqrt \pi \bar \Theta_2\right) \nonumber \\
\mu_5\sigma_4 &\sim& \sin \left(\sqrt \pi \bar \Theta_2\right).
\label{Isibosigmu45}
\eea
Since the $\xi^0$ field is decoupled from the interaction from
the beginning and the Majorana fermion $\zeta$ 
is massless at the 
Toulouse point, the Ising models noted ``0'' and ``4'' 
decouple from the other and remains critical. The correlation 
function of their order-disorder operators are therefore 
given by Eq. (\ref{Isigmucor}). However, the two Ising models
``3'' and ``5'' strongly interact with each other 
due to the hybridization of the Majorana $\xi^3$ and $\eta$ fields
at the Toulouse point (see Eq. (\ref{hybrid})).
As shown by the spectra (\ref{hybrispecp}, \ref{hybrispecm}), these two 
Ising models are non-critical and the order-disorder operators 
have short-ranged correlation functions characterized 
by the mass $m$ of the massive modes at the Toulouse point.

\section*{C Hybridized Ising models}

The aim of this appendix is to describe more precisely 
the consequence of the hybridization between the  
Majorana fermions $\xi^3$ and $\eta$ on their corresponding 
order-disorder ($\sigma_{3,5}, \mu_{3,5}$) Ising operators.
In particular, we shall prove, in the following, that
at the Toulouse point,  one has 
\bea 
\langle \mu_3 \mu_5 \rangle =  \langle \sigma_3 \sigma_5 \rangle
= \mu \ne 0 \nonumber \\
\langle \mu_3 \sigma_5 \rangle = \langle \sigma_3 \mu_5 \rangle
= 0.
\label{proffappend}
\eea

Let us begin by recalling the effective Hamiltonian
associated with the massive modes at the Toulouse point:
\bea 
{\cal H}_{hyb} =  - \frac{\ri v_1}{2}\left( \xi_R^{3}\partial_x \xi_R^{3}
-\xi_L^{3}\partial_x \xi_L^{3}\right)
- \frac{\ri u_2}{2}
 \left(\eta_R\partial_x \eta_R - \eta_L\partial_x \eta_L\right) 
\nonumber \\
+
 \ri m \left(\xi_R^{3}\eta_L - \eta_R\xi_L^{3}\right).
\label{hhybrappen}
\eea
For the computation of ground-state expectation values, 
we notice that the velocity anisotropy present in 
the Hamiltonian (\ref{hhybrappen}) can be ignored for 
several reasons.
On the one hand, we have two phenomenological parameters in
our approach: $v_1$ and $v_0$.
One can always choose $v_0 = v_1/2$ so that all spin velocities $u_1,u_2$
(\ref{velo}) of the modes
remain positive and $v_1=u_2$.
We expect that the results 
for the vacuum expectation values 
(\ref{proffappend}) obtained with this particular 
fine-tuning represent the generic situation.
On the other hand, one can justify further this proposition
by studying the structure of the equal-time fermionic Green's functions
of the Majorana fermions.
The nice thing is that the equal-time Green's functions
corresponding to the Hamiltonian (\ref{hhybrappen})
are in fact equal to that of a model with 
a single spin velocity (the average spin velocity):
\bea
{\tilde {\cal H}}_{hyb} =-\frac{\ri u_{*}}{2}\left(
\xi_R^{3}\partial_x \xi_R^{3}
-\xi_L^{3}\partial_x \xi_L^{3}\right)
 -\frac{\ri u_{*}}{2} \left(\eta_R\partial_x \eta_R
- \eta_L\partial_x \eta_L\right) \nonumber \\
+ \ri m \left( \xi_R^{3}\eta_L - \eta_R\xi_L^{3}\right)
\label{HB}
\eea
where $u_{*}$ is the average of the velocities $v_1$ and 
$u_2$: $u_{*} = (v_1+u_2)/2 = (v_1+v_0)/3$. 
The very reason for this equivalence on the equal-time Green's functions
stems from the fact that both Hamiltonian (\ref{hhybrappen}, \ref{HB})
have the same eigenvectors although different spectra.
Therefore, there is a one-to-one correspondence between the static
ground-state correlation functions of the model (\ref{hhybrappen})
and that of the Hamiltonian (\ref{HB}).
The velocity anisotropy will be important when considering
dynamical properties.
Consequently, our problem 
is reduced
to the computation of the ground-state 
expectation values (\ref{proffappend}) 
with the Hamiltonian (\ref{HB}).
To this end, let us introduce a bosonic field ${\tilde \Phi}$ 
and its dual ${\tilde \Theta}$: 
\begin{eqnarray} 
 \frac{\xi^3_{R} + \ri \eta_{R}}{\sqrt 2} &=& 
\frac{1}{\sqrt{2 \pi a_0}} \; \re^{\ri\sqrt{4\pi} {\tilde \Phi}_R} \nonumber \\ 
\frac{\xi^3_{L} + \ri \eta_{L}}{\sqrt 2} &=& 
\frac{1}{\sqrt{2 \pi a_0}} \; \re^{- \ri\sqrt{4\pi} {\tilde \Phi}_L}.
\end{eqnarray}
Using the results (\ref{Isibosigmu}) of the Appendix B, 
the different products 
of order and disorder Ising operators
can be expressed in terms of the bosonic fields:
\bea
\mu_3\mu_5 &\sim& \cos \left(\sqrt \pi  {\tilde \Phi} \right) \nonumber \\
\sigma_3\sigma_5 &\sim& \sin \left(\sqrt \pi  {\tilde \Phi} \right) \nonumber \\
\sigma_3\mu_5 &\sim& \cos \left(\sqrt \pi  {\tilde \Theta}\right) \nonumber \\
\mu_3\sigma_5 &\sim& \sin \left(\sqrt \pi  {\tilde \Theta}\right).
\label{IsingB}
\eea
The Hamiltonian (\ref{HB}) can also be rewritten in the 
following bosonized form: 
\begin{equation}
{\tilde {\cal H}}_{hyb} =  \frac{u_{*}}{2} \left(\left(\partial_x 
{\tilde \Phi}\right)^2 
+ \left(\partial_x {\tilde \Theta} \right)^2 \right)
- \frac{m}{\pi a_0}  \sin\left(\sqrt{4\pi} {\tilde \Phi}\right).
\label{HBboson}
\ee
Using the substitution: 
\begin{eqnarray}
{\tilde \Phi} &\rightarrow& {\tilde \Phi} + \frac{\sqrt{\pi}}{4} \nonumber \\
{\tilde \Theta} &\rightarrow& {\tilde \Theta},
\label{subsi}
\end{eqnarray}
${\tilde {\cal H}}_{hyb}$ identifies with the conventional 
form of a sine-Gordon
model at $\beta^2 = 4\pi$:
\begin{equation}
{\tilde {\cal H}}_{hyb} =  \frac{u_{*}}{2} 
\left(\left(\partial_x {\tilde \Phi}\right)^2 
+ \left(\partial_x {\tilde \Theta} \right)^2 \right)
- \frac{m}{\pi a_0}  \cos \left(\sqrt{4\pi}{\tilde\Phi}\right).
\label{HBboson1}
\ee
In this model, the bosonic field ${\tilde \Phi}$ 
is locked such as $\langle {\tilde \Phi} 
\rangle = 0$ ($m>0$).
Consequently, we have $\langle \cos \left(\sqrt \pi {\tilde \Phi} \right)\rangle
\ne 0$ and 
$\langle \sin \left(\sqrt \pi{\tilde \Phi} \right)\rangle
= 0$ whereas 
the dual field exponents $\exp\left(\pm \ri \sqrt \pi{\tilde \Theta} \right)$ 
have zero vacuum expectation values. 
Using Eq. (\ref{IsingB}) and the  substitution (\ref{subsi}),
we thus obtain the above mentionned result (\ref{proffappend}): 
\bea
\langle \mu_3 \mu_5 \rangle = \langle \sigma_3 \sigma_5 \rangle
= \mu \ne 0 \nonumber \\
\langle \mu_3 \sigma_5 \rangle = \langle \sigma_3 \mu_5 \rangle
= 0.
\label{proffappendfin}
\eea

\section*{D Stability of the Toulouse point solution}

The consistency
of the Toulouse approach to describe 
the universal properties of the chiral fixed point is investigated 
in this Appendix.

We first begin by analysing the effect of the operators that we
have neglected from the beginning in the Toulouse point
approach.
These operators are
the current-current interaction of same chirality
and the in-chain marginal irrelevant contribution
that appears in the continuum limit of each
S=1/2 Heisenberg spin chain (see Eq. (\ref{hcin0})):
\bea
{\cal O}_{cc} &=& {\bf J}_{0R} \cdot \left({\bf J}_{1R} +
{\bf J}_{2R} \right) + R \rightarrow L \nonumber \\
{\cal O}_{ic} &=& {\bf J}_{0R} \cdot {\bf J}_{0L} +
+ {\bf J}_{1R} \cdot {\bf J}_{1L}
+ {\bf J}_{2R} \cdot {\bf J}_{2L}.
\label{currneglect}
\eea

Let us first consider the current-current interaction of
same chirality (${\cal O}_{cc}$).
Using the bosonization of
a SU(2)$_1$ spin current (\ref{boso1}),
we can express the operator (${\cal O}_{cc}$) in terms of
the bosonic fields associated with each chain of the model:
\bea
{\cal O}_{cc} = \frac{1}{\sqrt{2} \pi} \left(\partial_x \varphi_{R}
\partial_x \Phi_R + \partial_x \varphi_{L} \partial_x \Phi_L
\right)  \nonumber \\
+ \frac{1}{2 \pi^2 a_0^2} \left(
\cos\left(\sqrt{8\pi} \varphi_{R} - \sqrt{4\pi} \Phi_R\right)
\cos\left(\sqrt{4\pi} \varphi_{-R}\right)
+ R\rightarrow L \right).
\label{curchirboso}
\eea
Using the canonical transformation (\ref{can1},
\ref{can2}), the current-current operator
can then be written as a function of the different fields of the
Toulouse point solution:
\bea
{\cal O}_{cc} = \frac{1}{\sqrt{2} \pi} \left(3 \partial_x {\bar
\Phi}_{2R} \partial_x {\bar
\Phi}_{1R} + 3 \partial_x {\bar
\Phi}_{2L} \partial_x {\bar
\Phi}_{1L}  - 2 \sqrt{2} \partial_x {\bar \Phi}_{1L}
\partial_x {\bar \Phi}_{1R} -
 2 \sqrt{2} \partial_x {\bar \Phi}_{2L}
\partial_x {\bar \Phi}_{2R} \right) \nonumber \\
+ \frac{1}{2 \pi^2 a_0^2} \left(
\cos\left(\sqrt{4\pi} \varphi_{-R}\right)
\cos\left(\sqrt{8\pi} {\bar \Phi}_1\right)
\cos\left(\sqrt{16\pi} {\bar \Phi}_{2R} +
\sqrt{4\pi}{\bar \Phi}_{2L}\right) \right. \nonumber \\
\left.
+ \cos\left(\sqrt{4\pi} \varphi_{-R}\right)
\sin\left(\sqrt{8\pi} {\bar \Phi}_1\right)
\sin\left(\sqrt{16\pi} {\bar \Phi}_{2R} +
\sqrt{4\pi}{\bar \Phi}_{2L}\right) +
R\rightarrow L \right) .
\label{curchirtous}
\eea
The next step of the calculation is to extract the
leading contribution of this operator in terms of
the different critical fields at the chiral fixed point.
Using Eqs. (\ref{psi}, \ref{cosphim}),
the non-zero expectation values (\ref{expec1}, \ref{expec2}),
 and the hybridization (\ref{hybrid})
of the massive fields, the leading part of the current-current operator of
same chirality
(\ref{currneglect}) at the chiral fixed point is given by:
\be
{\cal O}_{cc} \sim -\frac{2}{\pi} \partial_x {\bar \Phi}_{1L}
\partial_x {\bar \Phi}_{1R} + \frac{\gamma_m {\bar \gamma}_m}{\pi^2
a_0^2} \cos\left(\sqrt{8\pi} {\bar \Phi}_1 \right)
\label{curchieff}
\ee
up to contributions that will give
renormalization of mass and spin velocities.
One can express the operator (\ref{curchieff}) in terms of
the physical spin current (${\vec {\cal J}}$)
associated with the bosonic field ${\bar \Phi}_1$ (see Eq. (\ref{bosophi1})):
\be
{\cal O}_{cc} \sim -4 \left( {\cal J}^z_{R} {\cal J}^z_{L}
+\frac{\gamma_m {\bar \gamma}_m}{2}\left({\cal J}^+_{R} {\cal J}^-_{L}
+ H.c.\right)\right).
\label{curchieffnin}
\ee
At the SU(2) point, one has
$\gamma_m {\bar \gamma}_m =1$ (see Appendix A) and
the expression (\ref{curchieffnin}) can be recasted in a full rotational
form:
\be
{\cal O}_{cc} \sim -4 {\vec {\cal J}}_{R} \cdot
{\vec {\cal J}}_{L}.
\label{curchieffin}
\ee
The operator (\ref{curchieffin}) coincides thus with
the usual marginal irrelevant term of the continuum
limit of the effective S=1/2 Heisenberg spin chain
corresponding to the bosonic field ${\bar \Phi}_1$.
The contribution (\ref{curchieffin}) gives  logarithmic corrections
in  the spin-spin correlation functions at the chiral fixed point.

The second operator that we have neglected in our Toulouse
solution is the marginally irrelevant in-chain current-current
interaction:
\bea
{\cal O}_{ic} = {\bf J}_{0R} \cdot {\bf J}_{0L} +
+ {\bf J}_{1R} \cdot {\bf J}_{1L}
+ {\bf J}_{2R} \cdot {\bf J}_{2L}.
\label{currneglectirr}
\eea
We proceed in the same way as for the previous operator by
first rewritting ${\cal O}_{ic}$ in terms of the different
bosonic fields of each chain:
\bea
{\cal O}_{ic} = \frac{1}{8\pi} \left(\left(\partial_x \varphi\right)^2
- \left(\partial_x \vartheta\right)^2\right)
+ \frac{1}{8\pi} \left(\left(\partial_x \Phi\right)^2
- \left(\partial_x \Theta\right)^2\right)
+ \frac{1}{8\pi} \left(\left(\partial_x \varphi_-\right)^2
- \left(\partial_x \vartheta_-\right)^2\right) \nonumber \\
-\frac{1}{4\pi^2 a_0^2} \cos\left(\sqrt{8\pi} \varphi\right)
-\frac{1}{2\pi^2 a_0^2} \cos\left(\sqrt{4\pi} \Phi\right)
\cos\left(\sqrt{4\pi} \varphi_-\right).
\label{curirrbos}
\eea
At the Toulouse point,
this operator takes the following form:
\bea
{\cal O}_{ic} &=& \frac{1}{8\pi} \left(\left(\partial_x \varphi_-\right)^2
- \left(\partial_x \vartheta_-\right)^2\right)
+ \frac{3}{8\pi} \left(\left(\partial_x {\bar \Phi}_1\right)^2
- \left(\partial_x {\bar \Theta}_1\right)^2\right)
+ \frac{3}{8\pi} \left(\left(\partial_x {\bar \Phi}_2\right)^2
- \left(\partial_x {\bar \Theta}_2\right)^2\right) \nonumber \\
&-& \frac{1}{\sqrt{2} \pi} \partial_x {\bar \Phi}_1
\partial_x {\bar \Phi}_2
-\frac{1}{\sqrt{2} \pi} \partial_x {\bar \Theta}_1
\partial_x {\bar \Theta}_2
-\frac{1}{4\pi^2 a_0^2} \cos\left(\sqrt{8\pi} {\bar \Phi}_1\right)
\cos\left(\sqrt{16\pi} {\bar \Phi}_2\right)
\nonumber \\
&-&\frac{1}{4\pi^2 a_0^2} \sin\left(\sqrt{8\pi} {\bar \Phi}_1\right)
\sin\left(\sqrt{16\pi} {\bar \Phi}_2\right)
-\frac{1}{2\pi^2 a_0^2} \cos\left(\sqrt{8\pi} {\bar \Phi}_1\right)
\cos\left(\sqrt{4\pi} {\bar \Phi}_2\right)
\cos\left(\sqrt{4\pi} \varphi_-\right)
\nonumber \\
&-&\frac{1}{2\pi^2 a_0^2} \sin\left(\sqrt{8\pi} {\bar \Phi}_1\right)
\sin\left(\sqrt{4\pi} {\bar \Phi}_2\right)
\cos\left(\sqrt{4\pi} \varphi_-\right).
\label{curirrtous}
\eea
Using Eqs. (\ref{psi}, \ref{cosphim}), and the non-zero
expectation values (\ref{expec1}, \ref{expec2}),
the leading contribution of the operator (${\cal O}_{ic}$)
is at the chiral fixed point up to short-ranged parts
and terms giving a renormalization of the spin velocities:
\bea
{\cal O}_{ic} \sim  \frac{3}{2\pi} \partial_x {\bar \Phi}_{1L}
\partial_x {\bar \Phi}_{1R} -
\frac{\left(\gamma_m^2 + 2 {\bar \gamma}_m^2\right)}{4\pi^2
a_0^2} \cos\left(\sqrt{8\pi} {\bar \Phi}_1\right)
\label{curinchira}
\eea
and at the SU(2) symmetric point when $\gamma_m^2 =
{\bar \gamma}_m^2 =1$ (see Appendix A), it can be
reduced to the simple following form:
\bea
{\cal O}_{ic} \sim 3
{\vec {\cal J}}_{R} \cdot
{\vec {\cal J}}_{L}.
\label{curinchsu2fin}
\eea
Since the original coupling constant of the in-chain current-current
interaction (${\cal O}_{ic}$) is negative,
the effective operator (\ref{curinchsu2fin}) gives additional
logarithm corrections of the spin-spin correlation at
the chiral fixed point.

The next check of the correctness of the Toulouse solution is
to investigate the effect of small deviation of the coupling
constant $g_{\parallel}$ from its Toulouse-point value:
$g_{\parallel}^{*} = 4\pi(v_0+v_1)/3$.
In that case, the Hamiltonian (\ref{hfin}) at the Toulouse
point picks up an extra term:
\be
{\cal H}_{cor} = \frac{\delta g_{\parallel}}{\sqrt{2} \pi}
\left(\partial_x \varphi_L \partial_x \Phi_R
+ \partial_x \varphi_R \partial_x \Phi_L\right)
\label{touscorr}
\ee
with $\delta g_{\parallel} =  g_{\parallel} - g_{\parallel}^{*}$.
Using the canonical transformation (\ref{can1}, \ref{can2}),
this operator transforms into:
\bea
{\cal H}_{cor} = -\frac{\delta g_{\parallel}}{\pi}
\left( \left(\partial_x {\bar \Phi}_{1L}\right)^2 +
\left(\partial_x {\bar \Phi}_{1R}\right)^2
+ \left(\partial_x {\bar \Phi}_{2L}\right)^2 +
\left(\partial_x {\bar \Phi}_{2R}\right)^2 \right) \nonumber \\
+ \frac{3\delta g_{\parallel}}{\sqrt{2} \pi}
\left(
\partial_x {\bar \Phi}_{1L} \partial_x {\bar \Phi}_{2R} +
\partial_x {\bar \Phi}_{1R} \partial_x {\bar \Phi}_{2L}\right).
\label{touscorrtous}
\eea
Neglecting the renormalization of the velocities $u_{1,2}$,
we end with using Eq. (\ref{parphi2rl}):
\bea
{\cal H}_{cor} \sim
 \frac{3\delta g_{\parallel} \sqrt{\pi}}{\sqrt{2}} \ri
\left(
\partial_x {\bar \Phi}_{1L} \zeta_R \eta_R +
\partial_x {\bar \Phi}_{1R} \zeta_L \eta_L\right).
\label{touscorrlead}
\eea
Since the field
$\eta$ is massive,  the expansion in $\delta g_{\parallel}$
does not introduce new IR singularities, implying that
the long-distance behaviour of the correlation functions
will not be modified except for velocities
and mass renormalization. Therefore,
we can conclude that
the solution at the Toulouse point captures all
universal properties of the
chiral fixed point.

\end{document}